\documentclass[11pt]{article}

\usepackage{amsmath}
\usepackage{amsfonts}
\usepackage{amssymb}
\usepackage{latexsym}

\usepackage{cite}

\renewcommand{\d}{\mathrm{d}}

\newcommand{\dd}[2][]{\ensuremath{\frac{\d #1}{\d #2}}}
\newcommand{\pp}[2][]{\ensuremath{\frac{\partial #1}{\partial #2}}}

\newcommand{\Ra}{\Rightarrow}

\newcommand{\dga}{{\dot{\alpha}}}

\newcommand{\ga}{\alpha}
\newcommand{\gb}{\beta}
\renewcommand{\gg}{\gamma}
\newcommand{\gd}{\delta}
\renewcommand{\ge}{\epsilon}

\newcommand{\gf}{\phi}

\newcommand{\gm}{\mu}

\newcommand{\gl}{\lambda}

\newcommand{\gth}{\theta}

\newcommand{\gs}{\sigma}

\newcommand{\gp}{\pi}
\newcommand{\gps}{\psi}
\newcommand{\get}{\eta}

\newcommand{\gG}{\Gamma}
\newcommand{\gD}{\Delta}
\newcommand{\gF}{\Phi}

\newcommand{\gL}{\Lambda}

\newcommand{\gTh}{\Theta}

\newcommand{\gP}{\Pi}
\newcommand{\gPs}{\Psi}

\newcommand{\cD}{{\cal D}}

\newcommand{\cN}{{\cal N}}

\newcommand{\cW}{{\cal W}}

\newcommand{\ua}{{\underline a}}
\newcommand{\ub}{{\underline b}}
\newcommand{\uc}{{\underline c}}

\newcommand{\tff}{{\widetilde f}}

\newcommand{\tk}{{\widetilde k}}

\newcommand{\tm}{{\widetilde m}}

\newcommand{\tv}{{\widetilde v}}
\newcommand{\tw}{{\widetilde w}}

\newcommand{\tG}{{\widetilde G}}

\newcommand{\tK}{{\widetilde K}}

\newcommand{\tW}{{\widetilde W}}

\newcommand{\Tr}{\mbox{Tr}}
\newcommand{\tr}{\text{tr}}

\newcommand{\Id}{\text{\small 1}\hspace{-3.5pt}\text{1}}

\newcommand{\ra}{\rightarrow}

\newcommand{\der}{\partial}

\newcommand{\inv}{^{-1}}
\newcommand{\dsp}{\displaystyle}

\newcommand{\ovr}[1]{{\overline{#1}}}

\newcommand{\labl}[1]{\label{#1}}
\newcommand{\Kh}{K\"{a}hler}
\newcommand{\beq}{\begin{equation}}
\newcommand{\eeq}{\end{equation}}
\newcommand{\barr}{\begin{array}}
\newcommand{\earr}{\end{array}}
\newcommand{\equ}[1]{\begin{gather} #1 \end{gather}}
\newcommand{\equa}[1]{\begin{align} #1 \end{align}}

\newcommand{\arry}[2]{\begin{array}{#1} #2 \end{array}}

\newcommand{\pmtrx}[1]{\begin{pmatrix} #1 \end{pmatrix}}

\newcommand{\non}{\nonumber}
\newcounter{oldcounter}

\newcommand{\bff}{{\bar f}}

\newcommand{\bw}{{\bar w}}

\newcommand{\bz}{{\bar z}}

\newcommand{\bC}{{\overline C}}
\newcommand{\bD}{{\overline D}}

\newcommand{\bJ}{{\overline J}}

\newcommand{\bL}{{\overline L}}
\newcommand{\bM}{{\overline M}}

\newcommand{\bU}{{\overline U}}

\newcommand{\bW}{{\overline W}}

\newcommand{\bgf}{{\bar\phi}}

\newcommand{\bgm}{{\bar\mu}}

\newcommand{\bgth}{{\bar\theta}}

\newcommand{\bgps}{{\bar\psi}}
\newcommand{\bget}{{\bar\eta}}

\newcommand{\bgF}{{\overline\Phi}}

\newcommand{\bgTh}{{\overline\Theta}}

\newcommand{\bgPs}{{\overline\Psi}}

\newcommand{\ba}[2]{\[\begin{array}{#2}\label{#1}}
\newcommand{\ea}{\end{array}\]}
\newcommand{\be}{\begin{equation}}
\newcommand{\ee}{\end{equation}}
\newcommand{\bea}{\begin{eqnarray}}
\newcommand{\eea}{\end{eqnarray}}

\newcommand{\MSbar}{$\overline{\mbox{MS}}$}

\begin{document}

\thispagestyle{empty}

\begin{flushright}
HD-THEP-06-32 \\ 
IC/2006/118 \\
SIAS-CMTP-06-8 \\ 
hep-th/0612092
\end{flushright}
\vskip 2 cm
\begin{center}
{\Large {\bf 
Effective Action of Softly Broken Supersymmetric Theories 
}
}
\\[0pt]
\vspace{1.23cm}
{\large
{\bf Stefan Groot Nibbelink$^{a,}$\footnote{
{{ {\ {\ {\ E-mail: grootnib@thphys.uni-heidelberg.de}}}}}}}, 
{\bf Tino S. Nyawelo$^{b,}$\footnote{
{{ {\ {\ {\ E-mail: tnyawelo@ictp.it}}}}}}},
\bigskip }\\[0pt]
\vspace{0.23cm}
${}^a$ {\it 
Institut f\"ur Theoretische Physik, Universit\"at Heidelberg, \\ 
Philosophenweg 16 und 19,  D-69120 Heidelberg, Germany
 \\[1ex]
Shanghai Institute for Advanced Study, 
University of Science and Technology of China,\\ 
99 Xiupu Rd, Pudong, Shanghai 201315, P.R.\ China
 \\}
\vspace{0.23cm}
${}^b$ {\it 
The Abdus Salam ICTP, Strada Costiera 11, 
I--34014 Trieste, Italy 
\\[1ex] 
 Department of Physics, University of Utah, \\ 
    115 S 1400 E, \# 201, 
    Salt Lake City, UT 84112-0830, USA
\\}
\bigskip
\vspace{1.4cm} 
\end{center}
\subsection*{\centering Abstract}

We study the renormalization of (softly) broken supersymmetric
theories at the one loop level in detail. We perform this analysis in a
superspace approach in which the supersymmetry breaking interactions
are parameterized using spurion insertions. We comment on the
uniqueness of this parameterization. We compute the one loop
renormalization of such theories by calculating superspace vacuum
graphs with multiple spurion insertions. To preform this computation
efficiently we develop algebraic properties of spurion operators, that
naturally arise because the spurions are often surrounded by
superspace projection operators. Our results are general apart from
the restrictions that higher super covariant derivative terms and some
finite effects due to non--commutativity of superfield dependent mass
matrices are ignored. One of the soft potentials induces 
renormalization of the K\"ahler potential.

\newpage 


\setcounter{page}{1}

\section{Introduction and Summary} 
\label{sc:intro}

It is known for a long time that supersymmetric theories possess many
amazing ultra--violet (UV) properties. One of them being the absence of
quadratic divergences. It is this property that has led to the
development of the Minimal Supersymmetric Standard Model (MSSM). 
For renormalizable models the only exceptions to this mild UV behavior
are quadratically divergent $D$--terms~\cite{Fischler:1981zk}, but
since those only arise when there are also mixed gravitation--gauge
anomalies, they are not relevant for MSSM phenomenology. For string
model building on the other hand these $D$--terms are very useful to
reduce the rank of the gauge group. For non--renormalizable theories
one can expect more sources of non--logarithmically divergences.

Almost all of the parameters of the MSSM encode possible  ways
that supersymmetry is broken in a soft way, while preserving the gauge
symmetries of the MSSM. The soft parameters have essentially been 
classified in by Girardello and Grisaru~\cite{Girardello:1981wz}: They 
are either gaugino masses, Hermitian and complex mass matrices for the
complex scalars or three--linear scalar couplings. In fact, there
are some additional interactions that can be conditionally
soft~\cite{Harada:1981ug}, see also discussion
in~\cite{Weinberg:2000cr,Chung:2003fi}. These parameters
can either be introduced on the component level of the supersymmetric
theory, or described by spurion insertions, $\gth^2$ and
$\bgth{}^2$. The latter approach is very powerful because it leaves
most of the supersymmetric structure intact. A drawback of this
approach is that the spurion parameterization is not unique
\cite{Chang:1985qd}: A given soft supersymmetry breaking interaction
can be parameterized in various different ways. (We return to this
issue in more detail.) The renormalization group equations (RGE's) of
these soft parameters have been investigated up 
to the two loop level~\cite{Martin:1993zk,Yamada:1994id,Jack:1994kd,Jack:1998iy,Jack:1999ud,Jack:2004dv}.

Most of these works focused on the MSSM or renormalizable models in
general. However, when we see the MSSM as an effective description of
a more fundamental theory, it seems unnatural to only restrict the
attention to renormalizable models. String theory is often considered
as a possible candidate for the UV completion of the MSSM. The
effective low energy models that can be derived from String theory
always seem to be non--renormalizable, therefore it is important to
have a detailed analysis of quantum corrections of such effective
supersymmetric models as well.

For exact globally supersymmetric models in four dimensions
(renormalizable or not) there has been a lot of investigations to 
UV properties. Such a model with up to two derivatives is described by
a real \Kh\ potential $K$ and two holomorphic functions, the
superpotential $W$ and the gauge kinetic function $f_{IJ}$ of chiral
multiplet $\gf^a$\,. Because of holomorphicity the superpotential and
the gauge kinetic function are very much constraint.  This is
reflected in various non--renormalization
theorems~\cite{Grisaru:1979wc} for these functions and lead to
impressive results to all
order~\cite{Novikov:1982px,Jones:1983ip,Shifman:1986zi}. The 
works~\cite{Seiberg:1993vc,Seiberg:1994bz,Intriligator:1994jr} show
that even a lot of non--perturbative information can be
obtained.  In particular, in the certain $\cN=2$ theories the full
non--perturbative superpotential has been
computed~\cite{Seiberg:1994rs}.

The situation for the \Kh\ potential receives quantum corrections at
all orders in perturbation theory.  The one loop \Kh\ potential in
supersymmetric theories was first investigated
by~\cite{Buchbinder:1994iw} and subsequently computed by many groups
(see
e.g.~\cite{Buchbinder:1994df,Grisaru:1996ve,deWit:1996kc,Pickering:1996gt,Buchbinder:1999jw,Brignole:2000kg}),
using dimensional reduction the result\footnote{When non--Abelian
gauge interaction are included those results and
ours~\cite{Nibbelink:2005wc,Nyawelo:2006rw} differ, because the
computation has been performed in different gauges: We have used the
't~Hooft--Feynman gauge, while the other computations have been
performed in the Landau gauge. We have confirmed that this is a gauge
artifact by using a gauge fixing that can interpolate between these
two gauges.} reads~\cite{Nibbelink:2005wc}  
\equ{
K_{1L} ~=~ 
- \frac 1{16\gp^2} \, 
\tr_{\rm Ad}\, h\inv m_C^2 \Big( 2 - \ln \frac {h\inv m_C^2}{\bgm^2} \Big)
~+~ \frac 1{32\gp^2}\, 
\tr\, M_W^2 \Big( 2 - \ln \frac {M_W^2}{\bgm^2} \Big)
~,  
\labl{1LKh}
}
where $\bgm^2 \,=\, 4\gp e^{-\gg} \gm^2$ defines the \MSbar\
renormalization scale and $h_{IJ} \,=\, (f_{IJ}+\bff_{IJ})/2\,$. The
FP--ghost and superpotential mass matrices are defined by  
\equ{
(m_C^2)_{IJ} ~=~ 2\,  \bgf T_I G T_J \gf~, 
\qquad 
M_W^2 ~=~ 
 G\inv \bW {G\inv}{}^T W~,
\label{mC2mW2}
}
respectively. Here $T_I$ are the Hermitian generators of the gauge
group and $W\,$, $\bW$ and $G$ denotes the second holomorphic
derivative of the superpotential: $W \,=\, W_{,ab}\,$, the second
anti--holomorphic derivative $\bW^{\ua\,\ub} \,=\,
\bW{,}^{\ua\,\ub}\,$ and the second mixed 
derivative of the \Kh\ potential: $G^\ua{}_a \,=\,K{,}^{\ua}{}_{a}\,$,
respectively. The renormalization of the
\Kh\ potential has also been investigated at the two loop
level~\cite{Buchbinder:1996cy,Buchbinder:2000ve,Petrov:1999qh};
complete results including gauge corrections can be found
in~\cite{Nibbelink:2005wc}. These one and two loop results
in~\cite{Nibbelink:2005wc} were obtained by using that the vacuum 
graphs in superspace precisely correspond to the graphs needed to
compute the renormalization of the \Kh\ potential. This is an
efficient approach because the number of topologies of vacuum graphs
is limited, while computing diagrams with an arbitrary number of
external legs is very involved.

The basic aim of this paper is to investigate the effective action of
(softly) broken supersymmetric theories at the one loop level. 
To calculate the renormalization of softly broken supersymmetric
models, we choose to use similar method which were used to compute the
quantum corrections to the \Kh\ potential. In particular, to reduce
the number of diagrams we would like to compute only vacuum
supergraphs. In order to be able to do this, the use of spurion
supersymmetry breaking proves crucial: The full supersymmetric and the
supersymmetry breaking terms can then be represented as interactions
in superspace, and their renormalization can be performed
systematically by computing vacuum graphs in superspace only. As
observed above the parameterization of  supersymmetry breaking effects
is far from unique. However, using some holomorphicity restrictions we
give unique definitions for scalar functions $\tK\,$, $\tk$ and $\tW$.

The vacuum supergraphs, that give rise to the renormalization of soft
supersymmetry breaking action, involve infinite sums of spurion
insertions. To be able to manage this it is convenient to introduce
spurion operators. These spurion operators in fact arise naturally
because in the (quadratic) action the spurions find themselves
surrounded by either chiral or vector superfield projection
operators: For the chiral multiplets these projection operators are
hidden in the definition of chirality of the superfields, while the
full superspace representation for the vector multiplets explicitly
contains the vector superfield projector. These spurion operators
possess interesting algebraic properties because they can be shown to
generate two dimensional Clifford algebras. This observation greatly
simplifies computations. To avoid having to explicitly compute
infinite sums of vacuum supergraphs we derive some results for
computing logarithms of functional determinants. Combining these
results with the algebraic properties makes it conceptually
straightforward to compute the one loop renormalization of softly
broken supersymmetric theories, which is parameterized by the one loop
corrections of the functions $\tK$ and $\tk$.

Our results for the renormalization of (softly) broken supersymmetric
theories are general apart from the restrictions: We observe that for
non--renormalizable theories, there are additional supersymmetry
breaking interactions possible that cannot be parameterized by the
functions we have introduced. These interactions, that involve higher
super covariant derivatives, can also receive renormalization. Because
in this paper we ignore all possible superspace derivatives on the
chiral superfield background, our computations are blind to quantum
effects that generate such operators. In addition, we encountered one
technical problem: In general we consider theories with many chiral
multiplets, therefore one can encounter many field dependent mass
matrices. (We run into superpotential, complex and Hermitian scalar
mass matrices, to name a few.) In general these matrices do not all
commute, as a consequence even seemingly straightforward one 
loop scalar integrals can become difficult to compute exactly. For
this reason we have preformed approximations such that all effects due
to non--commutativity of mass matrices are finite and at least
proportional to one commutator of some of these mass matrices.

As the main part of the investigations in this paper is rather
technical, we have decided to structure the paper as follows: In next
section, section~\ref{sc:results}, we first give a concise definition
of the (soft) supersymmetry breaking functions, which we call soft
potentials, to avoid ambiguities that can arise when using the
spurion description of supersymmetry breaking. After we have
introduced the necessary notation, we give our results of one loop
computation of these soft potentials. In section~\ref{sc:illustrations} 
we give some applications and illustrations of our general results. We
find the conditions for softness of general non--renormalizable
theories. In subsection~\ref{sc:RenormKh} we give a simple example of
the Wess--Zumino model with supersymmetry breaking which induces 
renormalization of the \Kh\ potential, and discuss to what extend this
is related to non--softness of the theory. Finally we derive the soft
potentials for  Super Quantum Electrodynamics (SQED) with soft
breaking in subsection~\ref{sc:softSQED}. The remainder of the paper
is devoted to the details of the one loop computation of these soft
potentials. Section~\ref{sc:preparing} lays the technical foundations
for this: We first develop the properties of spurion operators in
subsection~\ref{sc:spurion}. Next the quadratic action is derived from
which all the one loop vacuum bubble supergraphs can be obtained. The
final subsection~\ref{sc:onelooptrln} gives the general expression for
these vacuum supergraphs with chiral superfield or vector superfields
running around in the loop. The actual computation of the one loop
renormalization of the soft potentials is performed in
section~\ref{sc:oneloopcomp}, relying heavily on the material 
developed in the preceding section. In subsection~\ref{sc:gaugecontr}
we first consider the gauge contributions to the renormalization of
these functions, because they are technically easier than the ones
that result from the chiral multiplets. Their contributions is
described in detail in subsection~\ref{sc:chiralcontr}. The paper is
concluded with two appendices that discuss various one loop integrals
we encountered. In appendix~\ref{sc:basicintegrals} the two basic
integrals are calculated using dimensional regularization in which all
our results will be expressed. Appendix~\ref{sc:complintegrals} is
devoted to three types of more complicated integrals that arise in
section~\ref{sc:oneloopcomp}.

\section*{Acknowledgments}

We would like to thank Borut Bajc and  Goran Senjanovic 
for discussions and useful comments. TN would thank the Institute for 
Theoretical Physics  at the University of Heidelberg for their  hospitality 
during his visit.

\section{Results for One Loop Soft Potentials} 
\label{sc:results}

We consider globally supersymmetric gauge theories with
arbitrary holomorphic superpotential $W(\gf)$ and gauge kinetic
function $f_{IJ}(\gf)$ and real \Kh\ potential $K(\bgf,\gf)$ of the
chiral superfields $\gf^a$ and their conjugates $\bgf_\ua$. We assume
that all these functions are gauge invariant. And in particular, the gauge
kinetic functions is proportional to the Killing metric. The vector
superfield $V = V^I T_I$ are contracted with Hermitean generators
$T_I$ of some (non--)Abelian group. The supersymmetric action for this
theory is given by 
\equ{
S_{\rm susy} ~=~ \dsp \frac 12 \int \d^8 z\, 
K\big(\bgf \,e^{2V}, \gf\big) 
\,+\, 
\int \d^6 z\, \Big\{
W(\gf) + \frac 14\, f_{IJ}(\gf)\,\cW^{I} \cW^J
\Big\}
\,+\, \text{h.c.}~,
\label{Acsusy}
}
where we use the full and chiral superspace measures, 
$\d^8 z \,=\, \d^4x \, \d^4\gth$ and 
$\d^6 z \,=\, \d^4x \, \d^2\gth\,$, respectively, 
and the Hermitean conjugation acts on all terms. In this action we
have introduced the {(non--)\-Abe\-lian} superfield strength   
\equ{
\cW^I{}_\ga ~=~ 
- \frac 18\, \bD{}^2 \Big(e^{-2V} D_\ga e^{2V} \Big)~. 
}
To include generic (soft) supersymmetry breaking interactions we extend
this theory by including the following soft action 
\equ{ \dsp 
S_{\rm soft} ~=~ \dsp \frac 12 \int\! \d^8 z\, \Big\{ 
\gth^2 \bgth^2\, \tK\big(\bgf \,e^{2V} \!\!, \gf\big) 
+ \gth^2\, \tk \big(\bgf \, e^{2V}\!\!, \gf\big)
\Big\}
+ \!\! \int \! \d^6 z\, \gth^2\, \Big\{
\tW(\gf) + \frac 14\, \tff_{IJ}(\gf)\,\cW^{I} \cW^J
\Big\}
+ \text{h.c.}
\label{Acsoft}
}
This soft action is essentially identical to the supersymmetric action
\eqref{Acsusy}, except for the appearance of the  spurions $\gth^2$
and $\bgth^2\,$. We may therefore refer to $\tK(\bgf, \gf)$ as the
soft (\Kh) potential, and to $\tW(\gf)$ and $\tff_{IJ}(\gf)$ soft
superpotential and soft gauge kinetic function, respectively. As their
supersymmetric analogs, these soft functions are all assumed to be
gauge invariant. The soft function $\tK$ can result in terms like the
Hermitean scalar masses, while the soft superpotential $\tW$ can give
rise to complex scalar masses and tri--linear scalar couplings. The
constant part of the $\tff$ gives rise to gaugino masses. Notice that
we also have introduce the soft potential $\tk$ in \eqref{Acsoft}
which does not have an supersymmetric analog.

This brings us to an important issue concerning the uniqueness of
presenting the soft action using spurion superfields $\gth^2$ and
$\bgth^2$ as is done in \eqref{Acsoft}. This issue has been discussed
before in the literature: To obtain a unique definition
ref.~\cite{Chang:1985qd} advocates to represent all soft breaking as
$D$--terms, while Yamada \cite{Yamada:1994id} gives spurion dependent
transformations to bring the (divergent) quantum corrections back to
its starting form. The ref.~\cite{Chang:1985qd}'s choice to represent
all supersymmetry breaking terms as \Kh\ terms works in general, but
for our purposes this classification is not fine enough. Because
Yamada was consider renormalizable theories, he was able to explicitly
construct these transformations. Since we also would like to consider
more complicated non--renormalizable models, we do not want to rely on
the existence nor explicit construction of such transformations. To
resolve these ambiguities in the definition of the soft action
\eqref{Acsoft} we note that the physics encoded in them is uniquely
defined by the component action after the auxiliary fields have been
eliminated. This mean one needs to compute the scalar potential $V_F$
from \eqref{Acsusy} and \eqref{Acsoft} 
\equ{
-V_F ~=~
\tK \,+\, \tW \,+\, \ovr{\tW} \,-\, 
\Big( \tk{}^{\ua} + \bW{}^{\ua}  \Big) \, G_\ua{}^a  \, 
\Big( \ovr{\tk}{}^{a} + W^{a}  \Big)~. 
\label{scalarpot} 
}
The last term includes the standard $F$ term in supersymmetric
models. In addition the chiral fermion masses depend on the functions
$\tk$ and $\ovr{\tk}\,$: 
\equ{
L_{\rm ferm\ mass} ~=~ 
-\, \frac 12 \Big( W + \ovr{\tk} \Big)_{;\,a\,b}\, \gps^a \gps^b 
\,-\, \frac 12 \Big( \bW + {\tk} \Big)_{;}{}^{\ua\,\ub}\, \bgps_\ua \bgps_\ub 
~, 
\label{fermmass} 
}
where the subscript $; ab$ denotes the second \Kh\ covariant
holomorphic derivative. From these expressions it is clear, that there
are many different spurion representations that give rise to the same
physics that is encoded in the scalar potential and the fermion mass
terms. This explains the existence of transformations like the ones
used by Yamada~\cite{Yamada:1994id}.

An additional complication is that there are more expressions, which 
one can write down, that can lead to scalar potentials and fermion
masses similar to the ones quoted above. For example, one can consider
the interaction  
\equ{
\int\d^8 z\, \gth^2\bgth^2\, B D^2 A~,
\label{hscd}
}
with $A$ and $B$ arbitrary functions of the chiral multiplets and
their conjugates. It is not difficult to see that this in general
gives both modification of the scalar potential \eqref{scalarpot} and
the fermion mass term \eqref{fermmass}. If $B$ would be anti--chiral,
one can partially integrate the $D^2$ to the $\gth^2$, and show that
this can be absorbed into the function $\tk$. But in general such
terms can not be absorbed into the functions we already defined. 
Since there is no obvious symmetry forbidding such interactions,
one can expect that in general at the quantum level they will be
generated. Because these terms are generated by diagrams with more
super covariant derivatives on the external legs, the degree of
divergence will be less than other diagrams in which all super
covariant derivatives act on the superspace delta functions in the
internal of the loop of the supergraph. If the theory is indeed soft,
i.e.\ no quadratic divergences, therefore one expects that these terms
will be finite. Technically computing the quantum corrections of such
terms results in similar difficulties as computing higher derivative
corrections to supersymmetric theories. In this work we ignore the
possibility of generating these interactions, by assuming that 
the background of chiral superfields can be treated as strictly
constant.

A unique definition of these functions is obtained by rewrite the
action \eqref{Acsoft} as  
\equ{
S_{\rm soft} ~=~ 
\frac 12 \int\! \d^8 z\, 
\gth^2 \bgth^2\, 
\Big( \tK\,+\, \tW \,+\, \ovr{\tW} \Big) 
\,+\, \int  \d^6 \bz\,  \tk 
+ \text{h.c.}~. 
\label{Acsoft2}
}
From this we infer that we can define $\tK$ and $\tW$ uniquely by the
requirement that $\tK$ does not contain a sum of purely holomorphic
anti--holomorphic terms: The holomorphic part can be absorbed in
$\tW\,$. Moreover, because $\tk$ has appeared under the 
$\int \d^6\bz$ integral, if it is anti--holomorphic, it can simply be
absorbed into the anti--holomorphic superpotential $\bW$. On the other
hand, it can be absorbed in $\tK + \tW + \ovr{\tW}$ when its second
anti--holomophic derivative vanishes. A physical reason is that as
long as it does not posses a second anti--holomorphic derivative, it
does not contribute to the fermion mass term~\eqref{fermmass}, hence
only modifies the scalar potential~\eqref{scalarpot}: From that
expression one can read of the modifications of the
functions $\tK$ and $\tW$ that result in the same scalar potential. 
The holomorphicity constraints on the functions $\tK\,$, $\tk\,$ and
$\tW$ are  not respected by quantum corrections, as the calculations
below will show. Of course it is possible to preform the same
splitting on these corrections, we will not preform this explicitly
here, as this becomes rather cumbersome.

The one loop corrections to the effective soft potentials 
include divergent contributions to $1/\ge$ and finite parts. Even
though the aim of this work is to obtain finite contributions, the
divergent parts of the one loop soft potentials are useful as
well: They provide us with an important consistency check of our
results, because from them we can obtain (part of) the well--known one
beta functions  
\cite{Inoue:1982pi,Inoue:1983pp,Derendinger:1983bz,Gato:1984ya,Falck:1985aa} (for results including two loop beta functions see
\cite{Martin:1993zk,Yamada:1994id}) for the soft parameters, if we
restrict to renormalizable models. We obtain these beta functions by
computing the renormalization of the parameters that appear in the
scalar \eqref{scalarpot}, and found exact agreement.

In the remainder of this section we give the results of analysis. First
of all the superpotential mass $M_W^2$ defined in~\eqref{mC2mW2} is
modified to 
\equ{
m_W^2 ~=~ \bw w~, 
\quad 
w_{ab} ~=~ W_{;ab} \,+\, \ovr{\tk}_{;ab}~, 
\quad 
(G w G^T)^{\ua\,\ub} ~=~ \bW_{,}{}^{\ua\,\ub} 
\,+\,\tk_{,}{}^{\ua\,\ub}~. 
\label{mW2mod}
}
Strictly speaking we only find normal, not covariant,
derivatives. This is a consequence of the fact that we assume that the
background satisfy the background equations of motions, which means
that various first derivatives are zero. The first effect of the one
loop renormalization of the renormalization of the supersymmetry
breaking action \eqref{Acsoft} is somewhat surprising: Because of the
modification of the superpotential mass matrix in \eqref{mW2mod} the
\Kh\ potential of the supersymmetric sector of the theory is modified to 
\equ{
K_{1L} ~=~ 
-\, \tr_{\rm Ad} L_1(m_C^2) 
\,+\,  \frac 12\,\tr\, L_1(m_W^2)~, 
\label{modKh1L}
}
where the integral $L_1$ is defined in
appendix~\ref{sc:basicintegrals}, 
which results from replacing $M_W^2$ by $m_W^2$ in \eqref{1LKh} and
modification of the counter term corresponding to the wavefunction
renormalization. The reason for this effect is that $\tk$ gives
similar terms as the anti--holomorphic superpotential $\bW$ as long as
only anti--holomorphic derivatives are applied.

In addition to the mass $m_W^2$ defined in \eqref{mW2mod}, we introduce
the mass matrices:   
\equ{
m_V^2 ~=~ \frac 12 \big(m_C^2 + m_C^2{}^T\big)~, 
~
m_G^2 ~=~ 2\,T_I \gf \, h\inv{}^{IJ}\, \bgf T_J G~, 
~
m_S^2 ~=~ - G\inv \tG~,  
~
\tm^2 ~=~ \big( \tw \ovr{\tw} \big)^{1/2}~, 
\label{mS2}
}
where have used the notations  
\equ{
\tG^\ua{}_a ~=~ \tK^\ua{}_a \,-\, 
\ovr{\tk}{}^{\ua}{}_{b}\, (G\inv)^b{}_{\ub}\, \tk{}^{\ub}{}_{a}~, 
\quad 
\tw_{ab} \,=\, \tW_{,\, ab} \,+\, \tK_{,\, ab}\,- \frac 12 \Big(  
w_{ac} \,(G\inv)^c{}_{\uc}\, \tk{}^\uc{}_b \,+\, a \leftrightarrow b 
\Big)~,
\label{extranot}
}
and $\ovr{\tw} \,=\,   G\inv {\tw}^{\dag} {G\inv}^T\,$. 
The first two masses are the vector and Goldstone boson masses in the
't Hooft--Feynman gauge. The latter two matrices are soft masses. 
These definitions can be intuitively understood.  However, we need to
explain why the definition of $\tG$ also includes the second mixed
derivatives of $\tk\,$, and why the expression for $\tw$ also contains
the second holomorphic derivative of $\tK$ and the second mixed
derivative of $\tk\,$: As explained above the definitions of the soft
functions is not unique. When defining the quadratic action (which is
used to obtain our results) similar ambiguities arise. We have
resolved them by making choices, that are most convenient for the
computation of the one loop corrections. 
Finally we define  mass matrices $m_\pm^2$  
\equ{
m_\pm^2 ~=~ 
h\inv m_V^2 + \frac 18\, |h\inv \tff|^2 \,\pm\, 
\sqrt{ \Big( h\inv m_V^2 + \frac 18\, |h\inv \tff|^2\Big)^2
\,-\,\big(h\inv m_V^2\big)^2 }~, 
\label{mpmgauge}
}
which involve the gaugino mass $\tff\,$.

We present the corrections to the effective one loop soft 
potentials in the following: It is natural to split the contributions
into terms that are proportional to $\gth^2$ (and their conjugates
proportional to $\bgth^2$), and terms proportional to
$\gth^2\bgth^2$. The detailed derivation of these results in the next
sections is preformed up to the level that we obtain standard
scalar integrals $J_0\,$, $J_1$ and $L_0\,$ defined in appendix
\ref{sc:basicintegrals}. In terms of these integrals we obtain for
$\tk$ at one loop 
\equ{
 \tk_{\rm 1L} ~=~ 
\frac 12 \int\limits_0^1 \! \d v\, 
\tr \Big[ \Big\{ 
J_0(m_{v+}^2) \,-\,  J_0(m_{v-}^2) 
\,+\, \frac 12\, m_S^2 J_1(m_{v+}^2) \,-\, \frac 12\, m_S^2 J_1(m_{v-}^2) 
\Big\} 
\tm^2 \frac 1{\ovr{\tw}}\, \bw
\Big]
\non \\[0ex] 
\,+\, \frac 12\, \gD R_0 \,-\, \frac 14\, \gD R_1 
\,+\, \frac 1{2(D-1)} \, \tr_{\rm Ad} \, \Big[
 \frac 1{m_+ -m_-} \, \Big\{
m_+\, J_{0}(m_+^2) \,-\, m_-\, J_{0}(m_-^2)
\Big\}\, 
h\inv \tff\, \Big]~.
\label{tk1Lint}
}
This result combines the expressions given in \eqref{chiralgth} and
\eqref{gaugegth} on the first and second line, respectively. We have
defined the mass matrices 
\(
m_{v\pm}^2 \,=\, m_S^2 \,\pm\, v\, \tm^2 \,+\, v^2\, m_W^2\,, 
\)
that depend on the integration variable $v\,$.
For $\tK$ we find 
\equa{\dsp 
 \tK_{\rm 1L} ~=\, & \dsp 
\tr \Big[
L_0(m_G^2) 
\,-\,   L_0(m_G^2+m_S^2)
\,+\, L_0(m_S^2)  
\,-\, \frac 12\, L_0(m_S^2+m_W^2 + \tm^2) 
\,-\,  \frac 12\, L_0(m_S^2+m_W^2 - \tm^2) 
\non \\[2ex] & \dsp  
\,+\, L_0(m_W^2)  
\Big]  
\,-\, \frac 12\, \gD K 
\,+\, \tr_{\rm Ad} \Big[
L_0(m_+^2) \,+\, L_0(m_-^2) \,-\, 2\, L_0(h\inv m_V^2)
\Big]~. 
\label{tK1Lint} 
} 
Here we have collected the contributions given in
\eqref{chiralgthbgth_A}, \eqref{chiralgthbgth_get}, 
\eqref{chiralgthbgth_perp} and \eqref{gaugegthbgth}. In these
expressions $\gD R_0\,$, $\gD R_1$ and $\gD K$ represent further
corrections that only arise if the various mass matrices do not
commute. In appendix \ref{sc:complintegrals} we have defined these
functions such that they give only finite contributions and are
proportional to at least one commutator of such matrices. In principle
one might also expect that the gauge corrections result in similar
corrections which are proportional to commutators containing
$\tff\,$. However, because we assume that $\tff$ is gauge invariant
and proportional to the Killing metric, all such commutators vanish.

These expressions of these soft potentials can now be evaluated 
using dimensional regularization. The standard integrals $J_0\,$,
$J_1$ and $L_0$ are evaluated in appendix \ref{sc:basicintegrals}, see
\eqref{J011/2exp} and \eqref{L0andL1exp}. In this scheme the finite
parts of these soft potentials is obtained by making the substitutions
\equ{
J_1(m^2) ~\ra~ \bJ_1(m^2) ~=~ 
 \frac {1}{16 \pi^2} 
\Big[  
1 \,-\, \ln \frac{m^2}{\bgm^2}
\Big]~, 
\quad 
L_0(m^2) ~\ra~ \bL_0(m^2) ~=~ 
-\frac 12\, \frac {m^4}{16 \pi^2} \, 
\Big[  
\frac 32 \,-\, \ln \frac{m^2}{\bgm^2} 
\Big]~,
\label{FiniteParts}
}
and using that within dimensional regularization 
$J_0(m^2) \,=\, - m^2\, J_1(m^2)\,$. We do not give the expression
resulting from substituting \eqref{FiniteParts} in \eqref{tk1Lint} and
\eqref{tK1Lint}, because they are somewhat lengthy and not very
illuminating. Moreover, for simplicity we do not give expressions for
the additional functions $\gD K\,$, $\gD R_0$ and $\gD R_1$ here,
because they are difficult to compute explicitly. On the other hand,
we see that --at least in these expressions-- the existence of
quadratic divergences is independent of the functions $\tk$ and
$\tW\,$.

\section{Ilustrations and Examples} 
\label{sc:illustrations}

\subsection{Conditions for Softness}
\label{sc:softness}

Even though we call the spurion insertions in action \eqref{Acsoft}
soft, these interaction might still lead to quadratic divergences,
because we generically consider non--renormalizable
theories. Therefore, we will here derive some simple criterion to 
ensure that the interaction terms \eqref{Acsoft} are indeed soft.

In this paper we predominantly use dimensional reduction to regularize
our quantum corrections at one loop. For the purpose of classifying
quadratic and logarithmic divergences this scheme is less useful
because both divergences just result in a pole in $\ge\,$. To
perform the classification of the quadratic divergences we therefore
here resort to the cut--off scheme, in which the loop momentum is
integrated upto scale $\gL\,$. Since we are only interested in one
loop vacuum graphs, i.e.\ without external lines, there are no
ambiguities in the definition of the cut--off scheme. Since we have
represented all results as sums of three types of standard integrals
$J_0\,$, $J_1$ and $L_0\,$, we simply have to give the representation
of their divergent behavior in the cut--off scheme. They read 
\equ{ 
J_0^{\rm div}(m^2) = \frac 1{16\,\gp^2} \Big[ 
\gL^2 - m^2\, \ln \gL^2 
\Big]
~,~
J^{\rm div}_1(m^2) = \frac 1{16\,\gp^2} \, \ln \gL^2
~,~
L^{\rm div}_0(m^2) = \frac{m^2}{16\,\gp^2}\,  \Big[ 
\gL^2 - \frac 12\, m^2\, \ln \gL^2 
\Big]~. 
}
Using these expressions to determine the quadratically divergent parts
of \eqref{tk1Lint} and \eqref{tK1Lint} we find 
\equ{
\tk_{\rm 1L}^{\rm quad\, div} ~=~ 
\frac {\gL^2}{96\,\gp^2} \tr_{\rm Ad} h\inv \tff 
~, 
\qquad 
 \tK_{\rm 1L}^{\rm soft} ~=~  
\frac {\gL^2}{16\,\gp^2} \Big\{ 
 \frac 14\, \tr_{\rm Ad} |h\inv \tff|^2
 \,-\, \tr\, m_S^2
\Big\}~, 
}
respectively. From these expression we conclude that if $h\inv\tff$
and $m_S^2$ are constants they give rise to constant quadratic
divergent corrections to the soft potentials which are unobservable,
otherwise quadratic divergence arise. In particular, as $m_S^2$ 
is a function of $\tK$ and $\tk$, one can say that if these 
functions are non-trivial quadratic divergences will arise and
softness is lost.

\subsection{Renormalization of the \Kh\ Potential}
\label{sc:RenormKh}

In the previous section we showed that the soft potential $\tk$,
defined in~\eqref{Acsoft} can induce a modification of the expression
of the one loop \Kh\ potential. As the \Kh\ potential describes the 
supersymmetric part of the theory, this is a surprising result. 
We explain that this additional renormalization is generically
accompanied by quadratic divergences, but one can consider models
where those are absent.

Because our general results are rather involved, we would like to give
a simple example in which these effects arise. To this end we consider
the renormalizable Wess--Zumino model described by the \Kh\ and
superpotential 
\equ{
K ~=~ \bgF \gF~, 
\qquad 
W ~=~ \frac 12 \, m\, \gF^2 \,+\, \frac 16\, \gl\, \gF^3~. 
\label{WZexample}
} 
To make our illustration as simple as possible, we take the soft
potential $\tk$ non--vanishing
\equ{ 
\tk ~=~ \frac 12\, \ell \, \bgF^2 \gF~,  
\label{WZexampleSoft}
} 
Using \eqref{modKh1L} we find for the \Kh\ potential 
\equ{
K_{1L} ~=~ \frac 12\, \frac {m_W^2}{16\,\gp^2}
\Big[  \frac 1\ge \,+\, 2 \,-\, \ln \frac {m_W^2}{\bgm^2} \Big]~, 
\qquad 
m_W^2 ~=~ \big| m \,+\, \gl\, \gF \,+\, \ovr{\ell} \, \bgF \big|^2~. 
}
Notice that it is not possible to reproduce the mass $m_W^2$ from a
modified superpotential alone, because it contains terms with $\gF^2$
which cannot be obtain from a superpotential alone. We can not
remove $\tk$ by a field redefinition of the scalars, because their
kinetic terms are then always modified. Also a modified $\gth^2$
dependent transformation does not work because such a transformation
necessarily has to violate chirality constraints. In the previous
subsection we have seen that a non--trivial $\tk$ generically induces
quadratic divergences.

From this one might conclude that the renormalization due to
supersymmetry breaking interactions only arises if these interactions
are not soft.  As we will now illustrate this generic conclusion is
not always true: The quadratic divergences can be fine tuned away at
least up to the one loop level. In particular, if we choose 
\equ{ 
\tK ~=~ \frac 14\, |\ell|^2\, \gF^2 \bgF{}^2~, 
}
we find that $m_S^2$ vanishes identically, and there are no quadratic
divergences. Hence renormalization of the \Kh\ potential is possible
even in model which are soft.

\subsection{Softly Broken SQED}
\label{sc:softSQED}

In this section we consider SQED with renormalizable soft breaking as
a particular illustration of the results described in this work. The
supersymmetric part of SQED is given by 
\equ{
S_{\rm SQED} ~= ~\int \d^8 z\, 
\bgf_\pm e^{\pm 2V} \gf_\pm 
\,+\, \int \d^6 z\, \Big(   
m\, \gf_+ \gf_- 
\,+\, \frac 1{4g^2} \int \d^6 z\, \cW^2
\Big) 
~+~ \text{h.c.}~, 
}
where $m$ is the mass of the electron superfield and $g^2$ the gauge 
coupling.  The notation $\pm$ in the first term indicates that we sum
over the kinetic terms of $\gf_+$ and $\gf_-\,$. The Hermitian
conjugation only acts on the chiral superspace integral part of this
expression. The corresponding soft action reads  
\equ{
S^{\rm soft}_{\rm SQED} ~= ~ -\int \d^8 z\, \gth^2\bgth^2\, 
m_0^2\, \gf_\pm e^{\pm 2V} \bgf_\pm \,+\, 
\int \d^8 z\, \gth^2 \Big(   
M^2 \, \gf_+ \gf_- 
\,+\, \frac 1{2 g^2} m_g\, \cW^2
\Big) 
~+~ \text{h.c.}~, 
}
where $m_0^2$ is a real soft scalar mass, $M$ a complex scalar
mass, and $m_g$ is the gaugino mass. The factor in front of the
gaugino mass has been chosen such that the normalization of the
kinetic term of the gaugino is taken into account. In principle there
can be two real scalar masses; different ones in front of the first
(two) terms. However, when one assume the 
discrete symmetry  
\equ{
\gf_\pm ~\ra~ \gf_\mp~, 
\qquad  
V ~\ra~ -V~, 
\label{discrete} 
}
these two masses are necessarily equal. In any case since all the
masses $m\,$, $m_{0}^2$ and $M^2$ are constants, the chiral
multiplets only give rise to constant corrections to the soft
potentials $\tK$ and $\tk$ at one loop, and therefore do not give rise
to observable renormalization.

We do receive corrections due to the gauge interactions, because we
encounter field dependent mass  matrices 
\equ{
m_V^2 ~=~ 2 g^2\, \big( |\gf_+|^2 \,+\, |\gf_-|^2 \big)~, 
\quad
m_G^2 ~=~ m_V^2\, \gP~, 
\quad 
\gP ~=~ \frac 1{|\gf_+|^2 + |\gf_-|^2} 
\pmtrx{ \gf_+ \\ - \gf_-} \pmtrx{\bgf_+ & - \bgf_-}~, 
}
where $\gP$ is a projection operator. Notice that we have absorbed a
factor $h\inv \,=\, g^2$ into the definitions of the mass
$m_V^2\,$. Finally the masses $m_\pm^2$ take the form 
\equ{
m_\pm^2 ~=~ 
m_V^2 + \frac 12\, |m_g|^2 \,\pm\, 
\sqrt{ \Big( m_V^2 + \frac 12\, |m_g|^2\Big)^2
\,-\,m_V^4 }~. 
} 
Using the general expression \eqref{tK1Lint} we obtain for $\tK$ the
one loop expression 
\equ{
 \tK_{\rm 1L} ~=~ 
L_0(m_+^2) \,+\, L_0(m_-^2) 
\,-\,  L_0(m_V^2) \,-\, L_0(m_0+m_V^2) 
~=~ - 
\frac {1}{32\, \gp^2} \Big[
(m_0^2 + m_V^2)^2 \ln \Big( 1 + \frac {m_0^2}{m_V^2} \Big) 
\non \\[2ex] 
\,+\, \Big\{ 
(m_+^2 - m_-^2)^2 + m_V^4 - (m_0+m_V^2)^2 
\Big\} \Big( \frac 32 - \ln \frac {m_V^2}{\bgm^2} \Big) 
- \frac{m_+^4-m_-^4}2 \ln \frac{m_+^2}{m_-^2}
\Big]
~. 
} 
For the other soft potential $\tk$ we find using \eqref{tk1Lint} 
\equ{
\tk_{\rm 1L} ~=~  -\frac{m_g}{48\, \gp^2} 
\Big\{
(m_+^2 + m_-^2 + m_+m_-)
\Big( 
\frac 53 - \ln \frac {m_+m_-}{\bgm^2} \Big)  
\,-\, \frac 12\, \frac {m_+^3 + m_-^3}{m_+ - m_-} \, 
\ln \frac {m_+^2}{m_-^2}
\Big\}~.  
} 
A few observations about these results are in order: 
These one loop corrections respect the discrete symmetry
\eqref{discrete}, showing that (at least) up to the one loop level
this discrete symmetry is respected. Also we see that they do not
dependent on the complex scalar mass $M$ at all, and $\tk_{\rm 1L}$
also does not depend on $m_0\,$.

A consistency check on these expressions is obtained when one
considers the supersymmetric limit, i.e.\ the tree parameters $m_0$ and
$m_g$ tend to zero, all the soft quantum corrections vanish. This is
indeed the case as can be seen from the leading behavior of the soft
potentials in this limit: 
\equ{
 \tK_{\rm 1L} ~=~ 
- \frac {m_V^2}{16\, \gp^2} \Big[ 
|m_g|^2 \Big( 1 - 2\, \ln \frac{m_V^2}{\bgm^2} \Big) 
\,-\, m_0^2 \Big( 1 - \ln \frac {m_V^2}{\bgm^2} \Big)
\Big]~,~ 
\tk_{\rm 1L} ~=~ 
- \frac{m_g m_V^2}{16\, \gp^2} 
 \Big( 1 - \ln \frac {m_V^2}{\bgm^2} \Big)~, 
} 
when $m_0, m_g \,\ra\, 0\,$.

\section{Preparing for One Loop Computations} 
\label{sc:preparing}

In this section we lay the basis for our one loop computation of the
soft potentials. This calculation can be thought of as an
extension of our computation of the \Kh\
potential~\cite{Nibbelink:2005wc,Nyawelo:2006rw} to the case where
supersymmetry is softly broken.

The aim is to compute the effective soft potentials $\tK$ and
$\tk$, see \eqref{Acsoft}, using a background field method  at the one
loop level. This mean that we will encounter one loop vacuum graphs in
this work, i.e.\ determinants of the kinetic operators of the various
quantum fields in the theory. For this reason we want to determine
the quadratic action of the quantum superfields, in which the masses
are functions of the background chiral multiplets. In this calculation
it is crucial to distinguish whether the spurions find themselves
surrounded by super covariant derivatives or not. When this is the
case the resulting spurion operators have many interesting algebraic
properties that we develop in detail in subsection~\ref{sc:spurion}.
(The super propagators as given in~\cite{Helayel-Neto:1984iv} can be
obtained using this treatment of the spurion insertions.) We can make
use of these algebraic properties, because we assume, that the
background chiral superfields $\gf^a$ are constant; the spurion
operators only act on each other or on superspace delta functions. The
limitation of this  procedure is that we are not able to compute the
possibility of additional higher super covariant derivative terms,
like the ones giving in \eqref{hscd}. However, for the computation of
the one loop expressions of the soft potentials $\tK$ and $\tk$ this
procedure is sufficient. Using the spurion operators we compactly
represent the quadratic action including soft terms in
subsection~\ref{sc:quadrac}. To compute the effective action amounts
to evaluating various functional determinants, in the final subsection
we review a basic method to do this.

\subsection{Algebras of Spurions}
\label{sc:spurion}

To describe supersymmetry breaking terms in a way that takes most
advantage of the special properties of supersymmetric theories, we
work in superspace and use spurions $\gth^2$ and $\bgth^2$ to
parameterize supersymmetry breaking terms. In this subsection we want
to develop some algebraic properties of spurion operators, that arise
when spurions find themselves between super covariant derivatives.

Because $\gth^\ga$ and $\bgth_\dga$ are Grassmannian variables, we 
obviously have that $(\gth^2)^2 \,=\, (\bgth^2)^2 \,=\, 0\,$. However,
as we will see in the next subsection, when these spurions appear in the
action of chiral superfields, then they are surrounded by the chiral
projectors  
\equ{
P_+ ~=~ \frac{\bD{}^2 D^2}{16 \,\Box}~, 
\qquad 
P_- ~=~ \frac{D^2 \bD{}^2}{16 \,\Box}~, 
}
which leads to interesting algebraic structure. To show this in the most 
transparent way, we define the chiral spurion operators 
\equ{
\get_\pm ~=~ \Box^{\frac 12}\, P_\pm \, \gth^2 \, P_\pm~, 
\qquad 
\bget_\pm ~=~ \Box^{\frac 12}\, P_\pm \, \bgth^2 \, P_\pm~. 
}
First of all, we should realize that these objects are operators
rather than simple Grassmannian numbers, as the original spurions
$\gth^2$ and $\bgth^2$ are. By definition $\get_\pm$ are left and
right chiral operators from both side, so properties like  
\(
\get_+ \get_- \,=\, \get_+ \bget_- \,=\, 0\,,  
\)
follow immediately. Also the algebraic properties can be verified
easily 
\equ{
\get_+^2 ~=~ \bget_+^2 ~=~ 0~, 
\quad 
\get_+ \bget_+ \get_+ ~=~ \get_+~, 
\quad 
\bget_+ \get_+ \bget_+ ~=~ \bget_+~. 
}
(Because identical properties can be obtained for $\get_-$ and
$\bget_-\,$, we do not describe them here explicitly.) 
Moreover, the products $\get_+\bget_+$ and $\bget_+\get_+$ are
not equal, as can be seen by writing both expressions in terms of the
original spurions 
\equ{
\get_+ \bget_+ ~=~ 
\frac{\bD{}^2}{-4} \, \gth^2 \bgth^2 \,\frac{D^2}{-4}~, 
\qquad 
\bget_+ \get_+ ~=~ 
\Box\, P_+\, \gth^2 \bgth^2 \, P_+~. 
}
These properties imply that $\get_+$ and $\bget_+$ generate a Clifford
algebra: 
\(
\big\{\, \get_+\,,\, \bget_+ \,\big\} ~=~ \Id_{\get_+}\,,
\)
where the combination $\Id_{\get_+} = \get_+\bget_+ + \bget_+\get_+$
plays the role of the identity, because 
$\get_+ \Id_{\get_+} = \Id_{\get_+} \get_+ = \get_+\,$, and similarly
for $\bget_+\,$. This means that we can identify $\bget_+$ and
$\get_+$ with Pauli matrices $\gs_+$ and $\gs_-\,$, respectively.
And therefore, we can interpret the combination  
\equ{
A_{\get_+} ~=~ 
A_{11}\, \bget_+ \get_+ \,+\, 
A_{12}\, \bget_+ \,+\, 
A_{21}\, \get_+ \,+\, 
A_{22}\, \get_+ \bget_+~
\label{Aexpansion}
}
as a $2\times 2$ matrix 
\equ{
A_{\get_+} ~=~ 
\pmtrx{ A_{11} & A_{12} \\[1ex] A_{21} & A_{22} }_{\get_+}~,
\label{Amatrix}
}
where $A_{ij}$ are constant complex numbers. It is also
straightforward to confirm that the product  $A_{\get_+} A_{\get_+}$
computed, either using the expansion \eqref{Aexpansion} and the above
given properties, or the matrix expression \eqref{Amatrix}, leads to
the same result. Using the identity $\Id_{\get_+}$ we can define the
projection operator  
\equ{
\perp_+ ~=~ P_+ \,-\, \get_+\bget_+ \,-\, \bget_+\get_+~, 
}
which is perpendicular to any $A_{\get_+}\,$: From the
expansion \eqref{Aexpansion} it follows immediately that 
\(
\perp_+ \, A_{\get_+} \,=\, A_{\get_+}\perp_+ \,=\, 0\,. 
\)
This implies that 
\equ{
\Big(A_0 \, \perp_+ +\, A_{\get_+} \Big)^2 
~=~ 
(A_0)^2 \, \perp_+ +\, (A_{\get_+})^2 
~,
\label{Afullproduct}
}
where $A_0$ is a complex number, just like the other $A_{ij}\,$.

In the following sections we will often use this
identification of the expansion \eqref{Aexpansion} with the matrix
\eqref{Amatrix} and the multiplication property \eqref{Afullproduct}
to simplify computations.  This is possible, because all statements
made above, can be generalized to the case where $A_0$ and $A_{ij}$
are $N\times N$ matrices themselves rather than merely complex
numbers. In that case we can consider the trace $\tr$ over such
$N\times N$ matrices. We generalize this notion to the matrices
$A_{\get_+}$ in the following way: 
\equ{
\tr\, A_{\get_+} ~=~ 
\pmtrx{ \tr\, A_{11} & \tr\, A_{12} \\[1ex] \tr\, A_{21} & \tr\, A_{22} }_{\get_+}
=~ 
\tr\, A_{11}\, \bget_+ \get_+ \,+\, 
\tr\, A_{12}\, \bget_+ \,+\, 
\tr\, A_{21}\, \get_+ \,+\, 
\tr\, A_{22}\, \get_+ \bget_+~. 
\label{trgetmat}
} 
Moreover, we can define two notions of transposition: We denote the
transposition of the $N\times N$ matrices by $A_{ij}^T$, and use the
symbol $A_{\get_+}^t$ to refer to the transposition of the entries of
the $A_{\get_+}$ matrix. In particular, on a matrix \eqref{Amatrix} of
$N\times N$ matrices these two types of transposition act as 
\equ{
\pmtrx{ A_{11} & A_{12} \\[1ex] A_{21} & A_{22} }_{\get_+}^T~=~ 
\pmtrx{ A_{11}^T & A_{12}^T \\[1ex] A_{21}^T & A_{22}^T }_{\get_+}~,
\qquad 
\pmtrx{ A_{11} & A_{12} \\[1ex] A_{21} & A_{22} }_{\get_+}^t~=~ 
\pmtrx{ A_{11} & A_{21} \\[1ex] A_{12} & A_{22} }_{\get_+}~. 
\label{transgetmat}
}
These definitions imply that 
$\tr\, A_{\get_+}^T \,=\, \tr\, A_{\get_+}\,$, 
but 
$\tr\, A_{\get_+}^t \,\neq\, \tr\, A_{\get_+}\,$.

For the vector superfields we will also encounter a special combination
of spurions and super covariant derivatives, similar to what  we have
seen above for chiral superfields. We associate to the projector on
the vector superfield degrees of freedom, 
\equ{
P_V ~=~ \frac{D^\ga \bD{}^2 D_\ga}{-8\, \Box} 
 ~=~ \frac{\bD_\dga D{}^2 \bD{}^\dga}{-8\, \Box}~, 
}
the spurion operators 
\equ{
\get_V ~=~ \frac{D^\ga \, \gth^2\, \bD{}^2 D_\ga }{8 \,\Box^{1/2}}~, 
\qquad 
\bget_V ~=~ \frac{\bD_\dga \, \bgth^2\, D^2 \bD{}^\dga }{8 \,\Box^{1/2}}~.
\label{ExpgetV}
}
They have been defined such that they have very similar properties as
$\get_\pm$ and $\bget_\pm\,$. In particular, they satisfy  
\equ{
\get_V{}^2 ~=~ \bget_V{}^2 ~=~0~, 
\quad 
\get_V \bget_V \get_V ~=~ \get_V~, 
\quad 
\bget_V \get_V \bget_V ~=~ \bget_V~,
}
and $\get_V \bget_V \,\neq\, \bget_V\get_V$ because 
\equ{
\get_V \bget_V ~=~ - \frac {i}{2}\, 
\gs^m_{\ga\dga}\, D^\ga \get_+ \bget_+ \bD{}^\dga
\,\frac {\der_m}{\Box}~, 
\quad 
\bget_V \get_V ~=~ - \frac {i}{2}\, 
\gs^m_{\ga\dga}\, \bD{}^\dga \bget_- \get_- D{}^\ga
\,\frac {\der_m}{\Box}~.
\label{ExpgetVbV}
}
Notice that any product between these vector spurion operators and
$\get_\pm$, $\bget_\pm$ vanish. As in the case of these chiral
spurions, we can say that $\get_V$ and $\bget_V$ generate a Clifford
algebra, and we can make the identification between functions of
$\get_V$ and $\bget_V$ and $2 \times 2$ matrices: 
\equ{
A_V ~=~ 
A_{11}\, \bget_V \get_V \,+\, 
A_{12}\, \bget_V \,+\, 
A_{21}\, \get_V \,+\, 
A_{22}\, \get_V \bget_V 
~=~ 
\pmtrx{A_{11} & A_{12} \\[1ex] A_{21} & A_{22} }_V~.
\label{MatV}
}
This identification is consistent in the sense that the product
$A_V\,A_V$ computed as functions of $\get_V$ and $\bget_V$ or as
matrices lead to the same result. In particular, we have the identity
matrix  
\(
\Id_V ~=~  \get_V\bget_V \,+\, \bget_V\get_V\,.
\)
Finally, we can also define the projector 
\equ{
\perp_V ~=~ P_V \,-\, \get_V\bget_V \,-\, \bget_V\get_V~, 
}
which is perpendicular to $A_V\,$: 
$\perp_V A_V \,=\, A_V \perp_V \,=\,0\,$. 
Therefore as chiral spurion operators we find: 
\equ{
\Big(A_0 \, \perp_V \,+\, A_{V} \Big)^2 
~=~ 
(A_0)^2 \, \perp_V \,+\, (A_{V})^2
~. 
}
Also here we can generalize all properties to matrices $A_0$ and
$A_{ij}\,$. This completes our exposition of the algebraic properties
of the spurion operators $\get_+$ and $\get_V\,$. 

We close this subsection by giving some relations to simplify these
spurion operators when they are sandwiched between two superspace
delta functions $\gd_{21}$ as naturally happens in the evaluation one
loop supergraphs. When they are placed between two superspace delta  
functions $\gd_{21}$ and are integrated over the full double
superspace, then these spurion operators reduce to simple spurion
insertions. For expression with both spurion operators of a given type
inserted, we find 
\equa{\dsp 
\int\d^8z_{12}\, \gd_{21}\, [F\, \get_\pm \bget_\pm]_2\, \gd_{21}
~=&\,  \dsp 
\int\d^8z_{12}\, \gd_{21}\, [F\, \bget_\pm \get_\pm]_2\, \gd_{21}
~=~ 
\int\d^4x_{12} \d^4\gth\, \gth^2 \bgth{}^2\, 
\gd^4_{21} \,F_2\, \gd^4_{21}~,
\label{RedCC} 
\\[2ex] \dsp 
\int\d^8z_{12}\, \gd_{21}\, [F\, \get_V \bget_V]_2\, \gd_{21}
~=&\,  \dsp 
\int\d^8z_{12}\, \gd_{21}\, [F\, \bget_V \get_V]_2\, \gd_{21}
~=~ 
-2\, \int\d^4x_{12} \d^4\gth\, \gth^2 \bgth{}^2\, 
\gd^4_{21} \,F_2\, \gd^4_{21}~, 
\label{RedVV}
}
where $\gd_{21} \,=\, \gd^4(x_2-x_1) \, \gd^4(\gth_2-\gth_1)\,$ and
the subscript $1$ or $2$ on the square brackets $[\ldots]$ and a given
superspace operator $F$ denote that the corresponding expression is
defined in (superspace) coordinate system $1$ or $2\,$.  When only a
single spurion operator of a given type is inserted, we instead obtain  
\equa{ \dsp 
\int\d^8z_{12}\, \gd_{21}\, [F \get_\pm ]_2\, \gd_{21}
~=&\,  \dsp 
\int\d^4x_{12} \d^4\gth^2\, \gth^2\, 
\gd^4_{21} \,\Big[F \frac 1{\Box^{1/2}} \Big]_2\, \gd^4_{21}~, 
\label{RedC}
\\[2ex] \dsp 
\int\d^8z_{12}\, \gd_{21}\, [F \get_V]_2\, \gd_{21}
~=&\,  \dsp 
2\, \int\d^4x_{12} \d^4\gth^2\, \gth^2\, 
\gd^4_{21} \,\Big[F \frac 1{\Box^{1/2}}\Big]_2\, \gd^4_{21}~, 
\label{RedV}
}
and similar relations are obtained for the conjugate spurion operator
insertions.

\subsection{Quadratic Action}
\label{sc:quadrac}

To be able to compute the one loop effective soft potentials we
expand the chiral multiplets around a non--trivial background of
chiral multiplets. To implement this we perform the background quantum
splitting by the shift $\gf ~\ra~ \gf \,+\, \gF\,$, where $\gf$ refers
to the classical background and $\gF$ is the quantum
fluctuation. We assume that the classical background is strictly
constant throughout the computation. This assumption is sufficient for
our purposes because we are interested in the computation of the soft
potentials $\tK$ and $\tk$ which do not involve any super covariant
derivatives. Treating $V$ also as a quantum fluctuation we end up with
the following quadratic action 
\equ{
S_{\rm quadr} ~=~ \int \d^8 z\, \Big\{
\bgF \big( G \,+\, \tG\, \gth^2\bgth^2 \big) \gF 
\,+\, 2\, \bgf V \big( G \,+\, \tG\,\gth^2\bgth^2 \big) \gF
\,+\, 2\, \bgF \big( G \,+\, \tG\,\gth^2\bgth^2 \big) V\gf 
\,+\, \qquad
\label{Acquadr}
\\[2ex]  
\quad \,+\, 2\, \bgf V \big( G \,+\, \tG\,\gth^2\bgth^2 \big) V \gf
\Big\} 
\,+\, 
\int \d^6z\, \Big\{ 
\frac 12\, \gF^T \big( w \,+\, \tw\, \gth^2 \big) \gF 
\,+\, \frac 14\, \big(f_{IJ} \,+\, \tff_{IJ}\, \gth^2 \big) W^{I\ga} W^J{}_\ga
\Big\} 
\,+\, \text{h.c.}~, 
\non 
}
where $W^I{}_\ga \,=\, -\bD{}^2 D_\ga V^I/4\,$, and the Hermitian
conjugation only acts on the $\int \d^6 z$ integral. Furthermore, we
have used the notation introduced in \eqref{extranot} which arises for
the following reasons: The functions $\tK$ and $\tk$ can possess
non--vanishing second holomorphic derivatives, resulting in the
definitions of $w$ and part of $\tw\,$.  The last term of $\tw$
in~\eqref{extranot} results from first eliminating the auxiliary fields
in this quadratic theory and then only reintroducing them for the
modified superpotential $w\,$. In this notation we keep the dependence
on the background chiral multiplets $\gf$ implicit.

An important distinction is now made depending on whether the spurions
find themselves surrounded by super covariant derivatives are not. If
they are not surrounded by such derivatives, they can at most be
inserted a {\em single} time in a diagram, otherwise the expression
for the diagram vanishes identically. Because, the superpotential and
the gauge kinetic terms are governed by holomorphicity, the spurions
are naturally surrounded by chiral projectors, therefore, the only two
point interaction which are not surrounded by super covariant
derivatives in \eqref{Acquadr} are given by  
\equ{
S_{\rm single} ~=~ 2 \int \d^8z\, \gth^2 \bgth^2\, \Big\{
\bgf V\, \tG\, V\gf \,+\, \bgf V\, \tG\,\gF \,+\, 
\bgF\, \tG\, V\gf \Big\}~.
\label{Acsingle}
}
We explain  at the end of this subsection that these interactions do
not lead to quantum correction to the soft potentials.

In the remaining terms of \eqref{Acquadr} the gauge superfields appear
in the same way as in the supersymmetry preserving theory, hence we
can use the gauge fixing action \cite{Ovrut:1981wa} (see
also~\cite{Nibbelink:2005wc}) 
\equ{
S_{\rm G.F.} ~=~ - \int \d^8z\, \gTh^I \, h_{IJ} \, \bgTh^J~, 
\qquad 
\gTh^I ~=~ \sqrt 2\, \frac {\bD{}^2}{-4} 
\Big[ V^I + h\inv{}^{IJ}\, \frac 1{\Box} \bgF\, G\, T_J\gf \Big]
\label{AcGF}
}
as if supersymmetry is unbroken: This gauge fixing is uniquely defined
by requiring that the mixing between the chiral and vector quantum
superfields is absent at the quadratic level, and that the $V$ 
propagator does not contain any $D$'s or $\bD$'s. This implies that
the FP--ghost sector is the same as in the supersymmetric theory in
the Feynman--'t~Hooft gauge~\cite{Ovrut:1981wa,Nibbelink:2005wc}. 
Therefore, in this gauge the ghosts only give standard corrections to
the \Kh\ potential (which are part of the result in \eqref{1LKh}), but
not to the soft potentials. Adding this gauge fixing action to the
remaining terms in \eqref{Acquadr} (that do not appear in
\eqref{Acsingle}), show that all mixing between the vector and chiral
superfields disappear. This allows us to separately give the quadratic
actions for the chiral and vector multiplets. Written as a full
superspace integral, the action for the chiral superfields becomes
after gauge fixing \eqref{AcGF}  
\equa{ \dsp 
S_\gF ~=&\, \dsp  \int\d^8z\, \Big\{ 
\bgF G \Big( 
 P_+ \,-\, m_G^2\,
\frac 1{\Box}\, P_+ \,-\, m_S^2\, \frac 1{\Box}\, \bget_+ \get_+ 
\Big) \gF 
\non \\[2ex] &  \dsp 
\,+\, \frac 12\, \gF^T \Big( 
w\, P_- + \tw\, \frac 1{\Box^{1/2}}\, \get_- 
\Big) \frac{D^2}{-4 \, \Box}\, \gF 
\,+\, \frac 12\, \bgF G \Big( 
\bw\, P_+ + \ovr{\tw}\, \frac 1{\Box^{1/2}}\, \bget_+ 
\Big) \frac{\bD{}^2}{-4 \, \Box}\, (G \bgF)^T~. 
\label{AcgF2}
}
Here we have made use of the chiral spurion operators $\get_\pm$ and
$\bget_\pm$ defined in section \ref{sc:spurion}. Similarly, using the
spurions $\get_V$ and $\bget_V$ defined there, we can write the
quadratic vector superfield action after gauge fixing as 
\equ{
S_V ~=~ - \int\d^8z\, V^I \Big\{
h_{IJ}\, \Box \,-\, m_{V\,IJ}^2
\,-\, \frac 12\, \tff_{IJ}\, \Box^{\frac 12}\, \get_V 
\,-\, \frac 12\, \ovr{\tff}_{IJ}\, \Box^{\frac 12}\, \bget_V 
\Big\}V^J~. 
\label{AcV2}
}
The quadratic superfield actions \eqref{AcgF2} and \eqref{AcV2} will
be the starting points of the computation of the effective soft
potentials $\tK$ and $\tk$ in section \ref{sc:oneloopcomp}.

We close this subsection by explaining why the two point interactions 
given in \eqref{Acsingle} do not lead to corrections to the soft
potentials. As we already observed because these terms have spurions 
without having covariant derivatives or projectors surrounding them,
they can be inserted at most a single time in a diagram. At the one
loop level this means, that the only possible diagrams contain a single
insertion of these operators and a single propagator closing the
diagram. Now it is important that in the gauge \eqref{AcGF} there are
no propagators that interpolate between a vector multiplet $V$ and
chiral or anti--chiral multiplets, $\gF$ or $\bgF$ (as can be seen
from the quadratic actions \eqref{AcgF2} and \eqref{AcV2}), therefore
the last two interactions in \eqref{Acsingle} cannot give a
contribution at one loop. The first term in \eqref{Acsingle} also
gives a vanishing contribution: From \eqref{AcV2} we can determine the
full propagator with the spurion supersymmetry
breaking~\cite{Helayel-Neto:1984iv} 
\equ{
\gD_V ~=~ \Big[ 1 \,-\, 
\frac 12\, \frac{|\tff|^2\, \Box}{(h\, \Box - m_V^2)^2} \Big]\inv 
\Big\{
\frac 1{h\, \Box - m_V^2} \, \Id_V 
\,+\, \frac 12 \, \frac 1{(h\, \Box - m_V^2)^2} 
\Big( 
\tff\, \Box^{1/2}\, \get_V + \ovr{\tff} \, \Box^{1/2} \, \bget_V
\Big)
 \Big\}~.
}
Because there is no mixing between the chiral and the vector
multiplets, there is only one diagram we can consider in which this
propagator is closed on the first interaction in
\eqref{Acsingle}. However, all possible contributions vanish because
of the explicit spurion appearance: 
\equ{ 
\arry{rl}{ \dsp  
\gd_{21}\, [\gth^2\bgth^2\, F\, \get_V ]_2\, \gd_{21}
~=\, & \dsp 
\gd_{21}\, \Big[\gth^2 \bgth{}^2\, 
\frac{ D^\ga \,\gth^2 \,\bD^2 D_\ga}{8\, \Box}  
\Big]_2 \gd_{21}~, 
\\[2ex] \dsp 
\gd_{21}\, [\gth^2\bgth^2\, F\, \get_V \bget_V]_2\, \gd_{21}
~=\, & \dsp 
- \frac {i}{2}\, \gs^m_{\ga\dga}\, \gd_{21}\, 
\Big[
\gth^2 \bgth{}^2\, 
D^\ga \frac{\bD{}^2}{-4} \gth^2 \bgth^2 \frac {D^2}{-4}  
\bD{}^\dga
\,\frac {\der_m}{\Box}\Big]_2 \gd_{21}~, 
}
} 
see \eqref{ExpgetV} and \eqref{ExpgetVbV}, using similar notation as
in the reduction formulae \eqref{RedCC}--\eqref{RedV}. In both
equations we see, that there are not enough super covariant derivatives
hitting the last spurions and the delta--function on the right, hence
both expressions are zero. We have therefore shown, that the
interactions \eqref{Acsingle} do not give any contributions to the
computation of the soft potentials.

\subsection{One Loop Functional Determinants}
\label{sc:onelooptrln}

To determine the one loop effective soft  potentials we develop
some general formalism to compute the corresponding functional
determinants efficiently. We will preform the calculation for both
chiral and vector multiplets, and aim to arrive at formulae, that can
be used both in the standard supersymmetry preserving as well as the
soft supersymmetry breaking situations.

We begin with chiral multiplets $\gF^a$. Their effective action
at one loop is given by 
\equ{
i\, \gG_\gF ~=~ 
\int | \cD \gF |^2 \, e^{i \, S_\gF}\Big|_{\rm conn}~, 
\qquad 
S_\gF ~=~ \frac 12 \int\d^8z\, 
\pmtrx{ \gF^T & \bgF } K \pmtrx{\gF \\ \bgF^T}~,
\label{lnDetCh}
}
the connected part of the Gaussian path integral. The quadratic
operator $K \,=\, P \,+\, L$ is decomposed into a standard free part 
\equ{
P ~=~ \pmtrx{0 & P_- \\[1ex] P_+ & 0}~,  
}
and an arbitrary perturbation $L\,$. Depending on the precise form of
$L$ this might be a rather involved computation.

To facilitate this computation we introduce a continuous parameter 
$0 \leq \gl \leq 1$, and define $K_\gl \,=\, P \,+\,\gl\,L\,$. 
Because the Gaussian integral of the standard free quadratic operator
$P$ is absorbed in the definition of the path integral, we can infer, that 
\equ{
i \gG_\gF ~=~ \int\limits_0^1 \d \gl\, \dd[]{\gl} \, i \, \gG_{\gl}~, 
}
where $\gG_\gl$ is defined as $\gG_\gF$ in \eqref{lnDetCh} but with
the interpolating quadratic operator $K_\gl$ instead of $K\,$. The point
of this exercise is, that the $\gl$ derivative of $i\, \gG_\gl$ is
technically simpler to compute than the original functional
determinant. Indeed, we find 
\equ{
 \dd[]{\gl} \, i \, \gG_{\gl} ~=~ 
 \int | \cD \gF |^2 \, e^{i \, S_\gl}
\, 
 \frac i2 \int\d^8z\, 
\pmtrx{ \gF^T & \bgF } L \pmtrx{\gF \\ \bgF^T}
\, 
\Big|_{\rm conn}~,
\label{lnDetChddgl}
}
corresponds to a single one loop diagram in the theory defined by the
interpolating kinetic operator $K_\gl$ with a single insertion of the
two point interaction defined by $L\,$.

Therefore to compute this we only need to determine the propagator in
the interpolating theory. By rewriting the action of the interpolating
action as 
\equ{
S_\gl ~=~ \frac 12 \int\d^8z\, 
\pmtrx{ \gPs^T & \bgPs } 
\,\frac{{\nabla^2}^T}{-4}\, K_\gl \,\frac{{\nabla^2}}{-4}\,
\pmtrx{\gPs \\ \bgPs^T}~, 
}
using the field redefinitions 
\equ{
\pmtrx{\gF \\ \bgF^T}
~=~ 
\frac{{\nabla^2}}{-4}\, \pmtrx{\gPs \\ \bgPs^T}~, 
\qquad 
\frac{{\nabla^2}}{-4} ~=~ 
\pmtrx{ 0 & \frac{\bD{}^2}{-4} \\[1ex] \frac{D^2}{-4} & 0 }~,
}
where $\gPs(\bgPs)$ is (anti--)chiral, it is straightforward to
confirm, that the propagator can be cast into the form 
\equ{
\gD_\gl ~=~ 
\frac {\nabla^2}{-4\,\Box} \big( \Id \,+\, \gl\, P^T L \big)\inv P^T 
\frac {{\nabla^2}^T}{-4\,\Box}~. 
}
Using this propagator and \eqref{lnDetChddgl} we obtain 
\equ{
\dd[]{\gl} \, i \, \gG_{\gl} ~=~ - \frac 12 \int d^8 z_{12}\, 
\gd_{21} 
\,\Tr\,\big[ P^T L ( \Id \,+\, \gl\, P^TL)\inv \big]\, 
\gd_{21}~. 
}
The notation $\Tr$ refers to the trace over the chiral multiplets in
the real basis defined by the vector  $(\gF^T, \bgF)^T\,$. This
expression is easily integrated over the parameter $\gl\,$, so that
the final result compactly reads  
\equ{
i \, \gG_{\gF} ~=~ - \frac 12 \int d^8 z_{12}\, 
\gd_{21} \,\Tr\, \ln \big[ P^T K \big]\,  \gd_{21}~. 
\label{EffAcgF}
} 
Because the FP--ghost are described by anti--commuting chiral
superfields, their one loop effective action is identical to the one
above except that the overall sign is opposite.

For the vector multiplets we can perform a very similar analysis to
compute the functional determinant 
\equ{
i\, \gG_V ~=~ 
\int \cD V\, e^{i\, S_V}\Big|_{\rm conn}~, 
\qquad 
S_V ~=~ - \int \d^8z\, V^I H_{IJ} V^J~, 
}
and we obtain 
\equ{
i\, \gG_V ~=~ - \frac 12\, \int\d^8z_{12}\,
\gd_{21} \,\tr_{\rm{Ad}}\, \ln \big[ H \big]\,  \gd_{21}~,
\label{EffAcV}
}
where $\tr_{\rm{Ad}}$ denotes the trace in the adjoint
representation.

\section{Computation of the One Loop Soft Potentials}
\label{sc:oneloopcomp}

In this section we present the details of the one loop computation of
the soft potential using the material developed in the previous
section which leads to the results quoted in section~\ref{sc:results}. 
The computations are somewhat involved, especially for the chiral
multiplets. Therefore we first present in subsection
\ref{sc:gaugecontr} the contributions due to gauge multiplets, because
they are technically a little easier than the ones that are due to the
chiral multiplets. They are present in the second subsection. In the
way we have setup the calculations we not only compute the soft potentials but also parts of the supersymmetry preserving \Kh\ 
potential. The well--known result of the one loop \Kh\ potential we
quoted in \eqref{1LKh} in the introduction, and we will have used it
as one of our cross checks on our computations.

\subsection{One Loop Soft Potentials due to Gaugino Masses}
\label{sc:gaugecontr}

We compute the one loop corrections to the soft potentials due
to the gaugino mass insertions. To this end we employ~\eqref{EffAcV} 
of subsection~\ref{sc:onelooptrln} to the quadratic action of the vector
multiplets, given in \eqref{AcV2} of subsection~\ref{sc:quadrac}, from
which the matrix $H$ can be read off. Using the 
spurion operators $\get_V$, $\bget_V$ and the projector $\perp_V$
defined in subsection~\ref{sc:spurion}, we can decompose this matrix as 
\equ{
H ~=~ (h\, \Box \,-\, m_V^2)\, \perp_V \,+\, (h\, \Box \,-\, m_V^2) 
\, \big( \Id_V + Z_V\big)~, 
}
where the vector multiplet mass matrix is given in~\eqref{mS2}. 
The matrix $Z_V$ has only off--diagonal block entries
\equ{
Z_V ~=~ - \frac 12 \, (h \, \Box \,-\, m_V^2)\inv  
\pmtrx{0 & \ovr{\tff}\, \Box^{\frac 12} \\[1ex] \tff\,\Box^{\frac 12} & 0}_V~.
\label{MatZV}
}
Here we emphasize with the subscript $V$ on the matrix, that we are
using the matrix notation for the corresponding expression
in terms of the spurions $\get_V$ and $\bget_V$ defined in
\eqref{MatV}. With $H$ written in this form it follows directly from
standard properties of taking $\tr\ln$ of matrices, that the effective
action, given in \eqref{EffAcV}, becomes 
\equ{
i\, \gG_V ~=~ - \frac 12\, \int\d^8z_{12}\, \gd_{12}\, 
\Big\{
\tr_{\rm Ad}\ln \big[ h\, \Box \,-\, m_V^2 \big] \,+\,  
\tr_{\rm Ad}\ln \big[ \Id_V \,+\, Z_V \big]
\Big\}_2 \gd_{21}~. 
\label{EffAcVint}
}
The first term does not give a contribution because there will
be no super covariant derivatives acting on the superspace delta
functions, so a standard supergraph theorem says that the result
vanishes.

To evaluate the second term directly is still difficult, therefore we
will use the same method as in subsection~\ref{sc:onelooptrln} to
compute the effective actions in general: We introduce a ficticious
parameter $0 \leq \gl \leq 1\,$, and define $Z_\gl \,=\, \gl\, Z_V\,$. We
can differentiate the  expression of \eqref{EffAcVint}, in which we
substitute $Z_V \,\ra\,Z_\gl$, w.r.t.\ $\gl$ to obtain 
\equ{
\dd{\gl} i\, \gG_\gl ~=~ 
- \frac 12\, \int\d^8z_{12}\, \gd_{12}\, 
\tr_{\rm Ad} \Big[ 
\big( Z_V \,-\, \gl\, Z_V^2 \big) 
\big(\Id_V \,-\,\gl^2\, Z_V^2 \big)\inv 
\Big]_2\, \gd_{21}~. 
}
The reason for this rewriting is, that now the whole expression is
written in terms of $Z_V^2$, expect for only the first term in the
first factor, so that we can now easily split the effective action in
``even'' and ``odd'' contributions: 
\equa{
\dd{\gl} i\, \gG^{\rm even}_\gl ~=&\, 
\frac 12\, \gl\, \int\d^8z_{12}\, \gd_{12}\, 
\tr_{\rm Ad} \Big[ 
Z_V^2 \big(\Id_V \,-\,\gl^2\, Z_V^2 \big)\inv 
\Big]_2\, \gd_{21}~,
\label{EffAcVintE}
\\[2ex] 
\dd{\gl} i\, \gG^{\rm odd}_\gl ~=&\, 
- \frac 12\, \int\d^8z_{12}\, \gd_{12}\, 
\tr_{\rm Ad} \Big[ 
Z_V \big(\Id_V \,-\,\gl^2\, Z_V^2 \big)\inv 
\Big]_2\, \gd_{21}~.
\label{EffAcVintO} 
}
Because the matrix $Z_V$, given in \eqref{MatZV}, is block
off--diagonal it follows that its square is block diagonal. 
Therefore, the even (odd) contributions leads to (off--)diagonal
contributions. In other words, using the identification 
\eqref{MatV} only the odd contributions lead to terms proportional to 
$\get_V$ or $\bget_V$, while the even ones give rise to effects 
proportional to products of these two spurion operators.

Note that it is straightforward to integrate the even part
\eqref{EffAcVintE}, and we obtain the compact expression 
\equ{
\gG^{\rm even} ~=~ 
-\frac 14\, \int\d^8z_{12}\, \gd_{12}\, 
\tr_{\rm Ad}\ln \Big[ \Id_V \,-\, Z_V^2 
\Big]_2\, \gd_{21}~. 
}
Because the expression for the square of $Z_V$ is a matrix with two
blocks on the diagonal, which are build out of the same matrices but
in opposite order, we find that the trace of both these blocks give
the same contributions. Moreover, using the identity for reducing
vector spurion operators between superspace delta functions
\eqref{RedVV}, we can write this as 
\equ{
\gG^{\rm even} ~=~ \int \d^8z\, \gth^2\bgth{}^2\!
\int\frac{\d^Dp}{(2\gp)^D\gm^{D-4}}\, \tr_{\rm Ad} 
\ln\Big[
\Id \,+\, \frac 14 \,p^2\, \frac1{p^2+ h\inv m_V^2} 
h\inv \ovr{\tff} \frac1{p^2+ h\inv m_V^2} h\inv \tff 
\Big]~. 
}
If the matrices $h\inv m_V^2$ and $h\inv \tff$ do not commute, this
expression is rather difficult to evaluate exactly. However, because
throughout this work we have assumed that the classical action is
gauge invariant, and in particular, that both the gauge kinetic
function $f_{IJ}$ and its soft analog $\tff_{IJ}$ are proportional to
the Killing metric, it follows that they do commute. Therefore, from
now on we can simply assume that the matrices $h\inv m_V^2$ and 
$h\inv\tff$ have already been diagonalized simultaneously. In this
case this integral becomes a sum of standard integral $L_0$ defined in
\eqref{Ln}  
\equ{
\gG^{\rm even} ~=~ \int\d^8z\, \gth^2\bgth{}^2\, 
\tr_{\rm Ad} \Big\{
L_0(m_+^2) \,+\, L_0(m_-^2) \,-\, 2\, L_0(h\inv m_V^2)
\Big\}~, 
\label{gaugegthbgth}
}
where we have used the mass matrices $m_\pm^2$ given in 
\eqref{mpmgauge} and reduction formula \eqref{RedVV}.

For the odd part of the effective action due to the soft gauge kinetic
function we make the same simplifying assumption on the mass
matrices $h\inv m_V^2$ and $h\inv \tff\,$, so that are already
diagonalized the odd part of the effective action can be written as  
\equ{
\gG^{\rm odd} ~=~ \frac 12 \int  \d^8 z 
\int\limits_0^1\!\d \gl 
\int \frac{\d^Dp}{(2\gp)^D \gm^{D-4}} \tr_{\rm Ad} 
\Big[ 
\frac {p^2\,+\, h\inv m_V^2}
{(p^2+h\inv m_V^2)^2 \,+\, \frac 14\gl \,|h\inv \tff|^2\, p^2}
\, 
h\inv \tff\, \gth^2\Big]
 \,+\, \text{h.c.}~, 
}
making use of \eqref{RedV}. This double integral has defined in
\eqref{intO} and is evaluated in appendix~\ref{sc:Oint}. We
obtain  
\equ{
\gG^{\rm odd} ~=~ 
\frac 1{2(D-1)} \int \d^8z\, \tr_{\rm Ad} \, \Big[
 \frac 1{m_+ -m_-} \, \Big\{
m_+\, J_{0}(m_+^2) \,-\, m_-\, J_{0}(m_-^2)
\Big\}\, 
h\inv \tff\, \gth^2
 \,+\, \text{h.c.}
\Big]
~,
\label{gaugegth}
}
where the mass matrices $m_\pm^2$ are defined in \eqref{mpmgauge}. 
This completes the computation of the corrections to the soft 
potentials $\tK$ and $\tk$ due to gauge interactions.

\subsection{One Loop Soft Potentials due to Chiral Multiplets}
\label{sc:chiralcontr}

We now turn to computation of the effective soft action due to chiral
multiplets. From the quadratic action for the chiral multiplets after
the gauge fixing \eqref{AcgF2} we read off its kinetic operator $K\,$,
which can be used in the functional determinant calculation for chiral
multiplets discussed in section \eqref{sc:onelooptrln}. We can
represent it in the following block matrix form 
\equ{
P^T K ~=~ \pmtrx{A_+ & \bC \\[1ex] C & A_-}~, 
}
where the matrices $A_+$, $A_-$, $C$ and $\bC$ are given by  
\equ{
\arry{cc}{ \dsp 
A_+ ~=~ \Big( \Id\,-\, m_G^2 \, \frac 1\Box \Big) P_+ 
\,-\, m_S^2\, \frac 1{\Box} \, \bget_+ \get_+~, 
\quad & \dsp 
C ~=~ \Big( w\, P_- \,+\, \tw\,\frac 1{\Box^{1/2}}\, \get_- \Big) 
\frac{D^2}{-4\, \Box}~, 
\\[2ex] \dsp 
A_- ~=~ \Big( \Id \,-\, {m_G^2}^T\, \frac1{\Box} \Big) P_-
\,-\, {m_S^2}^T \, \frac 1{\Box}\, \get_- \bget_-~, 
\quad  & \dsp 
\bC ~=~ \Big( \bw \, P_+ \,+\, 
\ovr{\tw}\, \frac 1{\Box^{1/2}}\, \bget_+\Big)
 \frac{\bD{}^2}{-4\, \Box}~. 
}
}
Here we have used the notations introduced below \eqref{mW2mod} and in
\eqref{mS2}. Notice that $A_- \,=\, A_+^T\,$ and that  
$\bC \,=\, C^\dag\,$. Using the expression for the one loop effective
action for general chiral multiplets, eq.~\eqref{EffAcgF}, we obtain 
\equ{
i\, \gG_\gF ~=~ - \frac 12 \int\d^8 z_{12} \,\gd_{21}
\Big\{
\tr\ln A_+ \,+\, \tr\ln A_- \,+\, \Tr\ln \Big[ \Id + Z \Big]
\Big\} \gd_{21}~, 
~
Z ~=~ \pmtrx{0 & A_+\inv \bC \\[0ex] A_-\inv C & 0}~. 
\label{EffAcgF2}
}
Next we compute the different parts of this expression separately.

Notice that the contributions of $A_+$ and $A_-$ are the same, because
these matrices are each others transposed, and there is a trace
$\tr\,$ over the whole expression. These contributions are easily 
computed by realizing that we can write 
\equ{
A_+ ~=~ \big(P_+ \,-\, \bget_+ \get_+ \big) 
\Big( \Id \,-\, \frac{m_G^2}{\Box} \Big) 
\,+\, \bget_+ \get_+ 
\Big( \Id \,-\, \frac{m_G^2+m_S^2}{\Box}\Big)~.  
}
Using this expression it is easy to understand that the sum of the
contributions due to $A_+$ and $A_-$ gives 
\equ{
\gG_{A} ~=~ 
\int\d^8 z\, \tr \Big[ L_1(m_G^2) \,+\, \gth^2\bgth^2\, 
\big\{ L_0(m_G^2) \,-\,   L_0(m_G^2+m_S^2) \big\} 
\Big]~,
\label{chiralgthbgth_A}
}
where the integrals $L_0$ and $L_1$ are defined in \eqref{Ln} of
appendix~\ref{sc:basicintegrals}, and we have used \eqref{RedCC}. 
The first term in this expression can be combine with the contribution
of the FP--ghost to the effective action to 
\equ{
\Big(\gG_{A} \,+\, \gG_{\rm FP}\Big)_{\rm SUSY} ~=~ 
- \frac 1{16\gp^2}\, \int\d^8z\, \tr \Big[ m_G^2 
\Big( \frac 1\ge \,+\, 2 \,-\, \ln \frac {m_G^2}{\bgm^2}  \Big) 
\Big]~. 
}
Notice this reproduces the result of the effective \Kh\ potential in a
supersymmetric theory given in \eqref{1LKh} (using that 
$\tr\, m_G^{2n}\,=\, \tr_{\rm Ad} (h\inv m_C)^{2n}$), and can
therefore be ignored when computing the effective soft potentials.

To compute the contribution due to the last term in \eqref{EffAcgF2} is
more work. We proceed in a similar fashion as before: Introduce  an
extra parameter $0 \leq \gl \leq 1$ in front of the off--diagonal
terms (the ones which are proportional to $C$ and $\bC$), and then 
differentiate w.r.t.\ this parameter. Next, we split the contributions
into ones with even and odd powers of $Z$. Non of the odd powers of
$Z$ give a contribution: Because  
\equ{
Z^2 ~=~ 
\pmtrx{A_+\inv \bC A_-\inv C & 0 \\[1ex] 
0 &  A_-\inv C A_+\inv \bC}~, 
}
is block diagonal, it follows that odd powers of $Z$ are necessarily
off--diagonal in the real basis of the chiral multiplets. But the
trace $\Tr$ in this basis is the sum of the traces in the complex
basis of the block diagonal parts, hence the odd powers do not
contribute. The even part is again readily integrated: Because all the
contributions are block diagonal in the real basis, the trace $\Tr$ in
the real basis reduces to the traces in the complex basis 
\equ{
i\, \gG_C ~=~ - \frac 14 \int \d^8z_{12}\, \gd_{21}\, 
\tr \Big[
\ln \Big(P_+ \,-\, A_+\inv \bC A_-\inv C \Big) \,+\, 
\ln \Big(P_- \,-\, A_-\inv C A_+\inv \bC \Big)
\Big]\, \gd_{21}~. 
\label{gGCint}
}
To evaluate this further we use the matrix notation introduced section
\ref{sc:spurion} to write 
\equ{
P_- -A_-\inv C A_+\inv \bC  = \, 
\perp_- \! \Big( 1 - \frac {w \bw}{\Box} \Big) 
\,+\, 
\Id_{-} - E_-~, 
\label{Ccontr}
}
where 
\equ{
E_- ~=~ 
\pmtrx{\ga & \gb \\ \gg & \gd}_{\get_-}
~=~
\pmtrx{ 1 & 0 \\ 0 &  \frac{\Box}{\Box-{m_S^2}{}^T} }
\pmtrx{ 
w \frac 1{\Box -{m_S^2}} \bw 
&
w \frac 1{\Box-{m_S^2}} \ovr{\tw} \, \frac 1{\Box^{1/2}} 
\\[2ex] 
\tw \frac 1{\Box -{m_S^2}} \bw\, \frac 1{\Box^{1/2}}
&
\frac 1{\Box} 
\Big( w \bw - \tw \frac 1{\Box-{m_S^2}} \ovr{\tw} \Big) 
}_{\!\get_-}.  
\label{E-}
}
The matrix $E_-$ has similar properties as the matrix $A_{\get_+}$
described in subsection \eqref{sc:spurion}. (The matrix notation after
the first equal sign is introduced here for later use below.) A
similar expression for the first combination in \eqref{gGCint} in terms of 
\equ{
E_+ ~=~ \left\{
\pmtrx{ 0 & \frac {\Box - {m_S^2}{}^T}{\Box} \\ 1 & 0 }
\, E_- \, 
\pmtrx{ 0 & 1 \\ \frac {\Box}{\Box-m_S^2{}^T} & 0}
\right\}^{t\, T}~, 
\label{E+andE-} 
}
where we would like to reminde the reader that $T$ denotes
transposition in the complex matrix basis, while $t$ denotes
transposition in $\get_\pm$--matrix basis, see~\eqref{transgetmat}. 
These expressions suggest how we can proceed in  
calculating the soft potentials: We first compute the effect
of the matrix $(\ldots)_{\get_-}$ and its conjugate, and after that
concentrate on the contributions due to the terms proportional to
$\perp_-\,$.

The computation of the soft potentials due to the $\get_-$--matrix
\eqref{Ccontr} (and its conjugate) is rather involved, and we again
resort to introducing a continuous parameter $0 \,\leq\, \gl \,\leq\,1\,$ 
in front of the $\get_-$--matrix part of $A_+\inv \bC A_-\inv C$. This
gives 
\equ{
\dd{\gl} i\, \gG^{\get_\pm}_\gl ~=~ 
\frac 14 \int\d^8z_{12}\, \gd_{12}\, \tr\, F[E_\pm]_2 \, \gd_{21}~, 
}
with $F[E] \,=\, E(\Id - \gl\, E)\inv\,$. (For notational simplicity
we have not made the $\gl$ dependence explicit in $F$.) Using the
expressions for $E_-$ and $E_+$ in eqs.~\eqref{E-} and
\eqref{E+andE-}, after a somewhat lengthy manipulations with these
matrices, we can show that the sum of $\gG^{\get_\pm}_\gl$ is given by 
\equa{ \dsp 
\dd{\gl} i\, \gG^\get_\gl ~=&\, \dsp 
\frac 12 \int d^4x_{12}\, \d^4\gth\,\gd^4_{21}\, \tr \Big[
\Big( F_{11}(E_-) + F_{22}(E_-) \Big) \gth^2 \bgth{}^2 
 \,+\, 
\frac 12 \Big( \frac {\Box - {m_S^2}{}^T}{\Box} + 1 \Big) 
\frac 1{\Box^{1/2}} F_{21}(E_-)\,\gth^2 
\non \\[2ex] & \dsp 
\,+\, 
\frac 12 \Big( \frac {\Box}{\Box - {m_S^2}{}^T} + 1 \Big) 
\frac 1{\Box^{1/2}} F_{12}(E_-)\,\bgth^2 
\Big]_2 \,\gd^4_{21}~,
\label{get_contr} 
}
using the identities \eqref{RedCC} and \eqref{RedC}. This expression
shows that we can distinguish between contributions that go
proportional to $\gth^2\bgth{}^2$ and those that only involve either
$\gth^2$ or $\bgth{}^2$.

We first focus on the $\gth^2\bgth{}^2$ part. Denoting the components
of the matrix $E_-$ in the $\get_-$--basis by $\ga\,$, $\gb\,$,
$\gg\,$ and $\gd\,$, as given in \eqref{E-}, we can show after some
algebra that 
\equ{
\tr \Big[ F_{11}(E_-) + F_{22}(E_-) \Big] ~=~ 
- \dd{\gl} \tr \Big[ 
\ln ( 1 \,-\, \gl\, \ga) \,+\, 
\ln \Big( 
1 \,-\, \gl\, \gd \,-\,\gg \frac {\gl^2}{1\,-\,\gl\,\ga} \gb
\Big) 
\Big]~. 
}
This means that for this contribution it is now straightforward to
integrate over $\gl$. By noting that 
\equ{
1 \,-\, \gd \,-\,\gg \frac {1}{1\,-\, \ga} \gb ~=~ 
\frac 1{\Box - {m_S^2}{}^T} 
\Big\{
\Box - {m_S^2}^T - w\bw 
- \tw \frac 1{\Box -m_S^2 - m_W^2} \ovr{\tw}
\Big\}~, 
\label{combi}
}
it follows that the $\gth^2\bgth{}^2$ contribution takes the form 
\equ{
\gG^{\get}_{\gth^2\bgth^2} ~=~ \int \d^8 z\, \gth^2 \bgth{}^2\, \Big\{ 
\tr\Big[ L_0(m_S^2) \,-\,  L_0(m_S^2+m_W^2) \Big] \,-\, 
\frac 12 \, K(m_S^2+m_W^2, \tw)
\Big\}~. 
\label{chiralgthbgth_getpre}
}
The simple matrix valued logarithmic integral $L_0$ is defined in
\eqref{Ln}, while the more complicated logarithmic integral
$K(m_S^2+m_W^2,\tw)$ is given in \eqref{Kint}. The latter integral is
very difficult to evaluate exactly if the mass matrices $m_S^2+m_W^2$
and $\tw$ do not commute. As is shown in appendix \ref{sc:Kint} we can
approximate $K$ by $K_0$, given in \eqref{K0}: The error $\gD K$ is
finite and at least proportional to a single commutator of these mass
matrices. We obtain 
\equ{
\gG^{\get}_{\gth^2\bgth^2} ~=~ \int \d^8 z\, \gth^2 \bgth{}^2\,  
\tr\Big[ 
 L_0(m_S^2)  
- \frac 12\, L_0(m_S^2+m_W^2 + \tm^2) 
-  \frac 12\, L_0(m_S^2+m_W^2 - \tm^2)
\Big] 
\,-\, \frac 12 \, \gD K~, 
\label{chiralgthbgth_get}
}  
where the mass matrix $(\tm^2)^2 \,=\, \ovr{\tw} \tw$ has been introduced. 
Notice that the second term in \eqref{chiralgthbgth_getpre} is
canceled by a contributions from $K_0\,$.

The term proportional to $\perp_-$ can be treated
independently because of property \eqref{Afullproduct} and results in
an standard logarithmic integrals \eqref{Ln}. A similar contribution
comes from the second term in \eqref{gGCint}, therefore we find the
total contribution due to $\perp_\pm$ in \eqref{Ccontr} is given by 
\equ{
\gG_{\perp} ~=~ \int \d^8z \tr \Big[
\frac 12\, L_1(m_W^2) \,+\, L_0(m_W^2) \, \gth^2 \bgth{}^2
\Big]~. 
\label{chiralgthbgth_perp}
}
The first term here corresponds to the supersymmetry preserving
correction to the one loop \Kh\ potential due to the
superpotential. This coincides with previous results
\cite{Grisaru:1996ve,Brignole:2000kg,Nibbelink:2005wc}, 
except that the mass $m_W^2$ here is the modified superpotential
mass~\eqref{mW2mod} rather then the true supersymmetric mass $M_W^2$
given in~\eqref{mC2mW2}. This means that in general even the
supersymmetric \Kh\ potential is modified by the effect of the soft
potential $\tk$. Notice that the second term also only depend on
$m_W^2$, therefore in the supersymmetric limit it has to be canceled
by other corrections. These other contributions are provided by the
second and third terms in \eqref{chiralgthbgth_get}: In the limit of
vanishing soft parameters they precisely cancel $L_0(m_W^2)$.

The last contribution we encounter comes from the off--diagonal parts
$F_{21}$ and $F_{12}$ in \eqref{get_contr}. They in the end give rise
to $\gth^2$ and $\bgth^2$ contributions. As one can show that these
contributions are each others complex conjugates, we only describe the
computation of the $\gth^2$ part explicitly here. By substituting the
expression \eqref{E-} for $E_-$ we obtain
\equ{ \!\!\! 
\dd{\gl} i\, \gG^\get_{\gth^2\, \gl} = \frac 14 \int \d^4 x_{12}\, 
\d^4 \gth\, \gth^2\, \gd_{21}^4\, \tr \Big[ 
\Big( 
2 - \frac {m_S^2{}^T}{\Box}
\Big) 
\Big( 
1 - \gl\, \gd - \gg \frac {\gl^2}{1-\gl\, \ga}\gb
\Big)\inv 
\gg \frac 1{1-\gl\, \ga} \frac 1{\Box^{1/2}}
\Big]_2 \, \gd_{21}^4~. 
}
Using the definitions of $\ga, \ldots, \gd$ given in \eqref{E-} and
the expression \eqref{combi} extended to include $\gl$, we can put the
integrated contribution in the form 
\equ{
\gG^\get_{\gth^2} ~=~  
\frac 14 \int \d^4 x\, \d^4 \gth\, \gth^2\, \Big\{
2\, R_0(m_S^2, \tw, w) \,-\, 
R_1(m_S^2, \tw, w)
\Big\}~,
}
where the function $R_0$ and $R_1$ are defined in \eqref{Rn} in
appendix \ref{sc:Rint}. As for the $\gth^2\bgth^2$ contribution
computed above, the integrals are difficult to perform explicitly,
therefore we again split finite terms that are only proportional to
commutators of combination of mass matrices and denote them by $\gD
R_0$ and $\gD R_1\,$. Using the results for $R_0$ and $R_1$ given in
\eqref{Rnfinal}, we can write the results compactly like 
\equa{\dsp 
\gG^\get_{\gth^2} ~=& \,   \dsp 
\frac 14 \int \! \d^4 x\, \d^4 \gth\, \gth^2 \Bigg\{ 
\int\limits_0^1 \! \d v\, 
\tr \Big[ \Big\{ 
2\, J_0(m_{v+}^2) -  2\, J_0(m_{v-}^2) 
+ m_S^2 J_1(m_{v+}^2) - m_S^2 J_1(m_{v-}^2) 
\Big\} 
\tm^2 \frac 1{\ovr{\tw}}\, \bw
\Big] 
\non \\[0ex] & \dsp \, 
\,+\, 2\, \gD R_0(m_S^2, \tw, w) \,-\, 
\gD R_1(m_S^2, \tw, w)
\Bigg\} 
~.
\label{chiralgth}
}
This result is written with an extra $v$ integration. To perform this
integration is not conceptually difficult, but will make the
expression rather lengthy even if when the mass matrices reduce to
mere (complex) numbers. This completes the description of the details
of the one loop contributions to the soft potentials $\tK$ and
$\tk$ due to the gauge interactions. The final results are collected
in section~\ref{sc:results}.

\appendix 
\def\theequation{\thesection.\arabic{equation}} 

\setcounter{equation}{0}

\section{Basic One Loop Scalar Integrals}
\labl{sc:basicintegrals}

This appendix is devoted to the evaluation of the basic one loop scalar
integrals, which arise in the main text of this paper. We compute these
scalar integrals in the \MSbar\ scheme: We evaluate the integrals in  
$D ~=~ 4 - 2 \ge$ dimensions, and we introduce the renormalization
scale $\gm$ such that all $D$ dimensional integrals have the same mass
dimensions as their divergent four dimensional counter parts.

The first basic type one loop integral is given by 
\equ{
J_n(m^2) ~=~ \int \frac{\d^D p}{(2\pi)^D \gm^{D-4}} \, 
\frac 1{p^{2n}}\, \frac{1}{p^2 + m^2} 
~=~ \frac {(m^2)^{1-n}}{16 \pi^2} \, 
\Big( 4\gp \frac {\gm^2}{m^2}  \Big)^{2- \frac D2}
\, \frac 1{\gG(D/2)}\, \frac {\gp}{\sin \gp(D/2-n)}
~, 
\label{Jn}
}
for $n \,=\, 0,1\,$. In the applications in the main text we need to expand
this to the zeroth order in $\ge$ including the pole $1/\ge\,$. In
particular, we encounter  
\equ{
J_0(m^2) = -\frac {m^2}{16 \pi^2}  
\Big[  
\frac 1\ge + 1 - \ln \frac{m^2}{\bgm^2} 
\Big]~,
\qquad 
J_1(m^2) = \frac {1}{16 \pi^2} 
\Big[  
\frac 1\ge + 1 - \ln \frac{m^2}{\bgm^2}
\Big]
~. 
\labl{J011/2exp}
}

The second class of integrals we encounter frequently in this work are
integrals over a single logarithm
\equ{
L_n(m^2) ~=~ \int \frac{\d^D p}{(2\pi)^D \gm^{D-4}} \, 
\frac{1}{p^{2n}} \ln\Big({1 + \frac{m^2}{p^2}}\Big)~.
\label{Ln}
}
In fact, this class of integrals and the previous ones are related to
each other via differentiation w.r.t.\ the mass parameter
\equ{
\pp{m^2}\, L_n(m^2) ~=~ J_n(m^2)
\quad \Ra \quad 
L_n(m^2) ~=~ \frac {m^2}{D/2-n}\, J_n(m^2)~,
\label{Lnsol}
}
for $n \,=\, 0,1\,$. In the second equation we have directly
integrated the general result for $J_n(m^2)$ given in \eqref{Jn}. In
the main part of the paper we need the following two integral results 
\equ{
L_0(m^2) ~=~ -\frac 12\, \frac {m^4}{16 \pi^2} \, 
\Big[  
\frac 1\ge + \frac 32 - \ln \frac{m^2}{\bgm^2} 
\Big]~, 
\qquad 
L_1(m^2) ~=~ \frac {m^2}{16 \pi^2} \, 
\Big[  
\frac 1\ge + 2 - \ln \frac{m^2}{\bgm^2} 
\Big]~. 
\label{L0andL1exp}
}

\setcounter{equation}{0}

\section{Complicated One Loop Scalar Integrals}
\labl{sc:complintegrals}

In this appendix we collect the computation of a number of one loop
integrals, which are more difficult to obtain directly, so that we can
avoid having length digressions on their computations in the main
text.

\subsection{$\boldsymbol{O(m^2, M^2)}$ Integral}
\label{sc:Oint}

The first integral we consider here is defined by 
\equ{
O(m^2, M^2) ~=~ \int\limits_0^1 \d \gl 
\int \frac{\d^Dp}{(2\gp)^D\gm^{D-4}}\, 
\frac {p^2+M^2}{(p^2 + M^2)^2 \,+\, 2 \,\gl^2\, m^2 \, p^2}~. 
\label{intO} 
}
After changing the order of integration, the integral over the parameter
$\gl$ can be rewritten as  
\equ{
O(m^2, M^2) ~=~ 
\frac {\sqrt 2}{m}
 \int\limits_0^\infty 
\frac {\d p\, p^{D-1}}{(4\gp)^{D/2}\gG(D/2) \gm^{D-4}}
\int\limits_0^{\tv(p)} \frac {\d v}{1+v^2}
~, 
\qquad 
\tv(p) ~=~ \frac {\sqrt 2\, m\, p}{p^2 + M^2}~, 
}
using the change of coordinates 
$ \sqrt 2\, \gl\, mp \,=\, (p^2+M^2)\, v$
(keeping $p$ fixed). Notice that only the upper limit is $p$
dependent, hence after performing an integration by parts, and
rewriting the resulting expression as a $D$ dimensional integral
again, we obtain  
\equ{
O(m^2, M^2) ~=~ \frac 1{D-1} 
\int \frac{\d^D p} {(2\gp)^{D}\gm^{D-4}} \,
\frac {p^2 \,-\, M^2}{(p^2+M^2)^2\,+ \, 2 \, m^2\, p^2}~. 
}
Using the factorization 
\equ{
\frac {p^2\,-\, M^2}{(p^2+M^2)^2 \,+\, 2\, m^2\,p^2} ~=~ 
\frac 1{m_+ -m_-} \, \Big\{
\frac {m_+}{p^2\,+\,m_+^2} \,-\, \frac {m_-}{p^2\,+\, m_-^2}
\Big\}~, 
}
with 
\( 
m_\pm^2 \,=\, m^2+ M^2 \,\pm \, \sqrt{(m^2+M^2)^2 \,-\, (M^2)^2 }\,,  
\) 
it is not difficult to confirm that the integral can be written in
terms of the simple integral $J_{0}$ as 
\equ{
O(m^2, M^2) ~=~ \frac 1{D-1}\, \frac 1{m_+ -m_-} \, \Big\{
m_+\, J_{0}(m_+^2) \,-\, m_-\, J_{0}(m_-^2)
\Big\}~. 
}

\subsection{$\boldsymbol{K(m^2,M^2)}$ Integral}
\label{sc:Kint}

This appendix is devoted to the computation of the integral 
\equ{
K(m^2, M^2) ~=~ \int \frac{\d^Dp}{(2\gp)^D\gm^{D-4}} 
\,\tr\, \ln \Big[
\Id \,-\, 
\frac 1{p^2\,+\, m^2{}^T} \,M^2\, 
\frac 1{p^2\,+\, m^2} \,\bM^2\, 
\Big]~, 
\label{Kint} 
}
where $m^2{}^\dag \,=\, m^2$ is a Hermitian matrix, and 
$\bM^2 \,=\, M^2{}^\dag$ is the Hermitian conjugate of a complex
matrix $M^2\,$. The main difficulty of this integral is that these
matrices not necessarily commute. We will show that we can do the
integral exactly if the matrices commute, and that the effects due to
non--commutativity of these matrices only gives additional finite
contributions.

We first explain how to determine the maximally commuting contribution
of this integral. Because $m^2$ is Hermitian, it can be diagonalized
by a unitary transformation $U\inv\,=\,U^\dag\,$. We can define 
\equ{
m^2 ~=~ U\inv \, m_D^2\, U~, 
\qquad 
M^2 ~=~ U^T\, M_D^2 \, U~, 
\qquad 
\tm_D^2 ~=~ \big( M_D^2 \bM{}_D^2 \Big)^{1/2}~, 
}
where the matrix $m_D^2$ is diagonal, but $M_D^2$ is not necessarily
diagonal. (Of course, if also $M_D^2$ is diagonal it implies that the
original matrices $m^2$ and $M^2$ commuted.) The matrix $\tm_D^2$,
defined using the formal power series of the square root, is
introduced so that we can write 
\equa{\dsp 
K(m^2, M^2) ~=\, &  \dsp 
\int \frac{\d^Dp}{(2\gp)^D\gm^{D-4}} \,\tr\, \ln \Big[
\Big( \Id + \frac{m_D^2}{p^2} \Big)\inv 
\Big( \Id + \frac {m_D^2+ \tm_D^2}{p^2} \Big) 
\Big( \Id + \frac{m_D^2}{p^2} \Big)\inv 
\Big( \Id + \frac {m_D^2- \tm_D^2}{p^2} \Big) 
\non \\[2ex] & \dsp 
\,-\, \frac 1{p^2+m_D^2}
\Big\{
\tm_D^2\, \frac {m_D^2}{p^2(p^2+m_D^2)} \, \tm_D^2 
\,-\, M_D^2\,  \frac {m_D^2}{p^2(p^2+m_D^2)} \, \bM_D^2 
\Big\}
\Big]~. 
\label{KintD} 
}
We can now consider the expansion in the second term. The zeroth order
of this expansion we denote by $K_0(m^2, M^2)$ and the rest of the
expansion we call $\gD K(m^2, M^2)\,$. The integral $K_0(m^2, M^2)$
can be evaluated in terms of the logarithmic integral $L_0$ as 
\equ{
K_0(m^2, M^2) ~=~ \tr\Big[
L_0(m_D^2+\tm_D^2) \,+\, L_0(m_D^2-\tm_D^2) 
\,-\, 2\, L_0(m_D^2) 
\Big]~, 
}
because the $\tr \ln$ of a product is equal to the sum of the $\tr\ln$
of the factors. Because of the overall trace, we can rotate
back to the original basis in which $m^2$ is not necessarily
diagonal, we then obtain 
\equ{
K_0(m^2, M^2) ~=~ \tr\Big[
L_0(m^2+\tm^2) \,+\, L_0(m^2-\tm^2) 
\,-\, 2\, L_0(m^2) 
\Big]~, 
\label{K0} 
}
where we have defined $\tm^2 \,=\, ( M^2 \bM{}^2 )^{1/2}\,$.

To conclude we show that the remaining part $\gD K(m^2, M^2)$ is
finite. To see this we write
\equa{\dsp 
\gD K(m^2, M^2) ~=\, &  \dsp 
\int \frac{\d^Dp}{(2\gp)^D\gm^{D-4}} \,\tr\, \ln \Big[
\Id \,+\, 
\frac 1{p^2\,+\, m_D^2-\tm_D^2} 
\,(p^2\,+\,m_D^2)\, 
\frac 1{p^2\,+\, m_D^2+\tm_D^2}\, \times 
\non \\[2ex] & \dsp \qquad \times \, 
\Big\{
\tm_D^2\, \frac {m_D^2}{p^2(p^2+m_D^2)} \, \tm_D^2 
\,-\, M_D^2\,  \frac {m_D^2}{p^2(p^2+m_D^2)} \, \bM_D^2 
\Big\}
\Big]~. 
} 
It is now easy to confirm that expanding this expression to any order
gives an integral which converges. In particular to first order, we
see that the integrant scales as $1/(p^2)^3$ for large $p^2\,$, and
hence is convergent. 

\subsection{$\boldsymbol{R_n(m_S^2,M^2,w)}$ Integral}
\label{sc:Rint}

Let $m_S^2$ be a Hermitian matrix, and $M^2$ and $w$ complex
matrices. We define for $n \,=\, 0,1$ the integrals 
\equa{\dsp  
R_n(m_S^2,M^2,w) ~=\, &  \dsp 
\int\limits_0^1 \d \gl \int \frac{\d^Dp}{(2\gp)^D\gm^{D-4}}\, 
\tr\Big[
\Big( \frac {m_S^2}{p^2} \Big)^n 
\Big(
1 \,-\, \gl \, 
\frac 1{p^2+m_\gl^2{}^T} M^2\frac 1{p^2+m_\gl^2} \bM^2 
\Big)\inv \, \times 
\non \\[2ex] & \dsp 
\, \times \, \frac 1{p^2+m_\gl^2{}^T}
M^2
\frac 1{p^2+m_\gl^2}
\bw
\Big]~,
\qquad 
 m_\gl^2 \,=\, m_S^2 \,+\, \gl\, m_W^2~. 
\label{Rn} 
}
We can follow a similar strategy as in the previous subappendix: First
go to a basis in which the propagators are diagonal, then compute in
that basis and finally rotate back. Because $m_\gl^2$
is a Hermitian matrix, there exist unitary $U_\gl$ such that 
$m_{D\,\gl}^2 \,=\, U_\gl \, m_\gl^2 \, \bU_\gl$ is diagonal. We
define 
$m_S^2 \,=\, \bU_\gl \, m_{SD}^2\, U_\gl\,$, 
$M^2 \,=\, U_\gl^T \, M_{D\,\gl}^2\, U_\gl\,$ and 
$w \,=\, U_\gl^T \, w_{D\,\gl}\, U_\gl\,$. 
(We use notation, like $w_{D\,\gl}$ to indicate that $w$ is evaluated
in the basis in which $m_\gl^2$ is diagonal. But since this matrix and
its diagonalization depend on $\gl\,$, also $w$ in this basis is $\gl$
dependent.) We can then write
\equa{\dsp  
&R_n(m_S^2,M^2,w) ~=~   \dsp 
-\int\limits_0^1 \d \gl \int \frac{\d^Dp}{(2\gp)^D\gm^{D-4}}\, 
\tr\Big[
\Big( \frac {m_{SD}^2}{p^2} \Big)^n 
\Big\{
\gl \,-\, \big[ \tm_{D\,\gl}^{-2} (p^2 + m_{D\,\gl})^2\big]^2 
\non \\[2ex]  \dsp 
&\,+\,  \tm_{D\,\gl}^{-2} (p^2 + m_{D\,\gl}^2) 
\tm_{D\,\gl}^{-2} (p^2 + m_{D\,\gl}^2)
\,-\, \bM_{D\,\gl}^{-2} (p^2+m_{D\,\gl}^2) 
M_{D\,\gl}^{-2}(p^2+m_{D\,\gl}^2) 
\Big\}\inv 
\bM_{D\,\gl}^{-2} \bw_{D\,\gl}
\Big]~. 
\label{Rexp}
}
If the matrices commute the last two terms under the big inverse
cancel; when they do not commute, the difference must be proportional
to a commutator. Clearly, if we Taylor expand in this difference 
\equ{
\gD R_n (m_S^2,M^2,w) ~=~ 
R_n(m_S^2,M^2,w) \,-\, R^0_n(m_S^2,M^2,w)~. 
}
we obtain convergent integrals, except for the zeroth order. 
The zeroth order contribution of \eqref{Rexp} can be rewritten as 
\equ{
R^0_n(m_S^2,M^2,w) = \!\!
\int\limits_0^1 \frac{\d \gl}{2 \gl^{1/2}} 
\int \frac {\d^D p}{(2\gp)^D \gm^{D-4}} \tr\, \Big[ 
\Big( \frac {m_{SD}^2}{p^2} \Big)^n 
\Big(
\frac 1{p^2+m^2_{D\,\gl-}} - \frac 1{p^2+m^2_{D\,\gl+}} 
\Big) 
\tm_{D\,\gl}^2 \bM_{D\,\gl}^{-2} \bw_{D\,\gl}
\Big],
}
where 
\( 
m_{D\,\gl\pm}^2 \,=\, 
m_{D\,\gl}^2 \,\pm\, \gl^{1/2}\, \tm_{D\,\gl}^2\,.
\) 
Transforming back to the original basis, and making a change of
variables $x \,=\, \sqrt \gl\,$, we obtain
\equ{
R^0_n(m_S^2,M^2,w) ~=~ 
\int\limits_0^1 \d v \, 
\tr\, \Big[ 
m_S^{2n} 
\Big( 
J_n(m^2_{v-}) \,-\, J_n(m^2_{v+}) 
\Big) 
\tm^2 \frac 1{\bM^2} \bw
\Big]~,
\label{Rnfinal}
}
with $m_{v\pm}^2$ defined below \eqref{tk1Lint}. 

\bibliographystyle{paper}
{\small
\bibliography{paper}
}


\end{document}